\documentclass[sigconf,nonacm]{acmart}
\usepackage{booktabs}
\usepackage[utf8]{inputenc}
\usepackage[range-phrase=~--~,range-units=single]{siunitx}
\usepackage{subcaption}
\usepackage{balance}

\makeatletter
\def\author@bx@sep{0.1pc}
\makeatother

\DeclareSIUnit{\dBm}{dBm}

\AtBeginDocument{%
  \providecommand\BibTeX{{%
    \normalfont B\kern-0.5em{\scshape i\kern-0.25em b}\kern-0.8em\TeX}}}

\setcopyright{none}
\renewcommand{\footnotetextcopyrightpermission}[1]{}
\copyrightyear{2023}

\begin{document}

\title
[Evaluating the Security of Power Conversion Systems Against Electromagnetic Injection Attacks]
{Assault and Battery: Evaluating the Security of Power Conversion Systems Against Electromagnetic Injection Attacks}

\author{Marcell Szakály}
\email{marcell.szakaly@cs.ox.ac.uk}
\affiliation{%
  \institution{University of Oxford}
  \city{Oxford}
  \country{United Kingdom}}

\author{Sebastian Köhler}
\email{sebastian.kohler@cs.ox.ac.uk}
\affiliation{%
  \institution{University of Oxford}
  \city{Oxford}
  \country{United Kingdom}}

\author{Martin Strohmeier}
\email{martin.strohmeier@ar.admin.ch}
\affiliation{%
  \institution{armasuisse S+T}
  \city{Zürich}
  \country{Switzerland}}

\author{Ivan Martinovic}
\email{ivan.martinovic@cs.ox.ac.uk}
\affiliation{%
  \institution{University of Oxford}
  \city{Oxford}
  \country{United Kingdom}}

\renewcommand{\shortauthors}{M. Szakály et al.}

\begin{abstract}
    Many modern devices, including critical infrastructures, depend on the reliable operation of electrical power conversion systems.
    The small size and versatility of switched-mode power converters has resulted in their widespread adoption.
    Whereas transformer-based systems passively convert voltage, switched-mode converters feature an actively regulated feedback loop, which relies on accurate sensor measurements.
    Previous academic work has shown that many types of sensors are vulnerable to Intentional Electromagnetic Interference (IEMI) attacks, and it has been postulated that power converters, too, are affected.

    In this paper, we present the first detailed study on switched-mode power converters by targeting their voltage and current sensors through IEMI attacks.
    We present a theoretical framework for evaluating IEMI attacks against feedback-based power supplies in the general case.
    We experimentally validate our theoretical predictions by analyzing multiple AC-DC and DC-DC converters, automotive grade current sensors, and dedicated battery chargers, and demonstrate the systematic vulnerability of all examined categories under real-world conditions.
    Finally, we demonstrate that sensor attacks on power converters can cause permanent damage to Li-Ion batteries during the charging process.
\end{abstract}

\maketitle

\section{Introduction}
\label{sec:intro}

The transition to renewable electricity, battery storage technologies, and electrification of every day appliances is driving the demand for smaller and more efficient power converters.
These can come in many different sizes, from mobile phone chargers, to Electric Vehicle (EV) chargers and grid-scale power converters.
Thanks to the proliferation of cheap semiconductors, many new power converters are using a \textit{switched-mode} architecture due to their small size and excellent efficiency~\cite{smps-takeover}.

Failure of a power converter renders the entire device unusable, and can even cause damage: many electronic components can be destroyed by applying too much voltage to their input, and Lithium-based battery chemistries are able to catch fire as a result of overcharging~\cite{yi2022safety}.
Global news coverage of the ``exploding'' Samsung Galaxy Note 7 series~\cite{samsung-battery}, and discussion about EV fires~\cite{ev-fires} demonstrate significant and unsurprising public interest in the topic.
While in most of these cases the root cause of the fault is primarily mechanical, a faulty charger can also cause combustion.

Switched-mode power supplies are actively regulated devices, they monitor their output and adjust their operation in order to achieve the desired output voltage or current.
A large body of previous work has shown that sensors are vulnerable to physical-layer attacks~\cite{yan2020sok, giechaskiel2020taxonomy}, and it has been proposed that attacking the sensors used inside switched-mode power supplies can affect their output and risk damage to downstream systems~\cite{dayanikli2020electromagnetic}.

To the best of our knowledge, no prior work presents a detailed theoretical and experimental evaluation of hardware attacks on power conversion systems, instead showing only attacks against standalone voltage sensors and simulated evaluations \cite{dayanikli2020electromagnetic, attaianese2022effects}.
The current EV powertrain security literature focuses on software and protocol level security~\cite{ye2021cyber, sripad2017vulnerabilities}, rather than on fundamental hardware properties of the system.
In this paper, we overcome these limitations and present the first evaluation of hardware attacks against voltage and current sensors, as they are used in real-world systems such as EV charging stations.
Our proposed wireless attack uses Intentional Electromagnetic Interference (IEMI) and can be carried out using cheap off-the-shelf equipment, making the barrier to entry for an adversary low.
An antenna is placed near the target and a malicious RF signal is emitted at the target's resonant frequency, which couples to wires inside the device and interferes with its correct operation.
More specifically, the malicious signal targets the feedback circuitry in switched-mode power converters, causing the device to produce an incorrect voltage or current that can damage downstream systems or the converter itself.
Due to the volatile chemistry of batteries and the associated safety risk that can arise from overcharging, we are also investigating the impact of our attack on battery chargers and battery cells.\\

\noindent\textbf{Contributions}
\\In summary, we make the following contributions in this paper:
\begin{itemize}
    \item We develop a mathematical framework to predict and precisely calculate the impact of an attack on feedback loop-based regulation systems.
    
    \item We successfully demonstrate for the first time, real-world IEMI attacks against power converters and automotive grade current sensors.
    
    \item We show that these observations also transfer to voltage and current sensors as used in battery chargers.
    
    \item We prove experimentally that our attack can cause permanent damage to Li-Ion batteries in a realistic setting.
    
    \item Finally, we present potential countermeasures to such attacks and discuss their inherent limitations.
\end{itemize}

\section{Related Work}
\label{sec:related}

Academic literature has presented several examples highlighting the importance of sensor reading integrity~\cite{yan2020sok, giechaskiel2020taxonomy}, with attack signals ranging from acoustic waves~\cite{trippel2017walnut, yunmok2015rocking} over light~\cite{kohler2021they, sugawara2020light} to coupled and wireless RF signals.
In particular, signal injection into analogue sensor systems using IEMI is a well documented attack vector, which has been demonstrated experimentally against a variety of systems.
IEMI attacks have been used to inject intelligible signals into microphones~\cite{xu2021inaudible, kune2013ghost}, cameras~\cite{kohler2022signal, jiangglitchhiker} and even digital communications~\cite{dayanikli2022wireless}, and have also been used to inject DC offsets into systems such as temperature sensors~\cite{tu2019trick}, voltage and current sensors~\cite{dayanikli2020electromagnetic}, and output driving signals~\cite{dayanikli2020electromagnetic, mohammed2022towards}.

With the rising popularity of EVs and renewable energy, a large body of work in the engineering and chemistry literature has examined practical and safe engineering solutions for battery management systems~\cite{bms-overview}.
However, these works do not consider adversarial scenarios.
In the security literature, research focusing on the EV powertrain mostly considers software and protocol level attacks~\cite{ye2021cyber, sripad2017vulnerabilities}, with some research proposing power converter and IEMI attacks against the battery charging system~\cite{dayanikli2020electromagnetic}.

Two previous papers are closely related to this present work, as they study the effects of IEMI on power converters.
In \cite{dayanikli2020electromagnetic}, the authors study a battery charger by simulation, explaining how a small change in charger voltage can result in a large change in the charge current.
In their experiments, they use an inductively coupled attack source, by placing a toroidal magnet around the voltage sensor wires, which is unable to scale towards a remote attack.
Their experiments show that both a standalone voltage and current sensor can be affected, with a \SI{200}{\milli \watt} signal source at close proximity able to change a measured voltage from \SI{21}{\volt} to \SI{42}{\volt}.
The paper does not perform an end-to-end attack, where a voltage sensor is attacked while being used in the regulation process.
Our paper improves upon this work, by attacking sensors as they are located inside real power converters, during operation, to study the real-world challenges of such an attack and the practical impact on the output.
Furthermore, we use a radiating attack source, which allows attacks from a distance.

In \cite{attaianese2022effects}, the authors study by simulation the effect of low frequency signals on boost converters.
Their results show that an attack affecting the voltage sensor measurement can affect the output voltage of the device.
The low frequency attack leads to an increased ripple in the output, and can even lead to an increase in the output voltage.

\section{Threat Model}

\paragraph{Goals}
We assume an adversary with the goal to wirelessly impair the correct functioning of a power supply, using only intentional electromagnetic interference.
The attack can cause a temporary denial of service by lowering the voltage or shutting down the system, or in a more serious case, cause permanent damage by increasing the voltage and current.
In view of the importance of large battery systems, as used in electric vehicles or as storage for photovoltaic systems, we assume that the attacker has a particular interest in battery charging systems, with the aim to cause permanent damage.
The intentions for such an attack can vary, ranging from denial-of-service to blackmail.

\paragraph{Capabilities}
Given the low barrier of entry for EMI attacks, we consider an adversary who has the budget and knowledge of a motivated hobbyist.
In addition, access to only commercial off-the-shelf (COTS) equipment is required.
This includes a signal generator or software-defined radio capable of operating at the required frequency range, a suitable directional antenna with high gain and, if necessary, a suitable RF amplifier to increase the attack distance.
Due to the nature of the attack, physical proximity to the target is usually required.
Nevertheless, with additional resources, appropriate amplification enables attacks from a distance.

While access to target feedback improves attack success and reduces the number of parameters to be tuned, we assume that the adversary operates without any feedback in a black box setting.
However, unrestricted access to an identical copy of the target device is required in order to determine the resonant frequency at which the attack is most effective.

\section{Background}
\label{sec:bg}

In this section we present a brief technical overview of power converters and battery charging, while in the next section we present a mathematical model of IEMI and regulation systems, and make predictions for the experiments performed.

\subsection{Power Converters}
\label{sec:bg_converter}

\begin{figure}
  \centering
  \includegraphics[width=\linewidth]{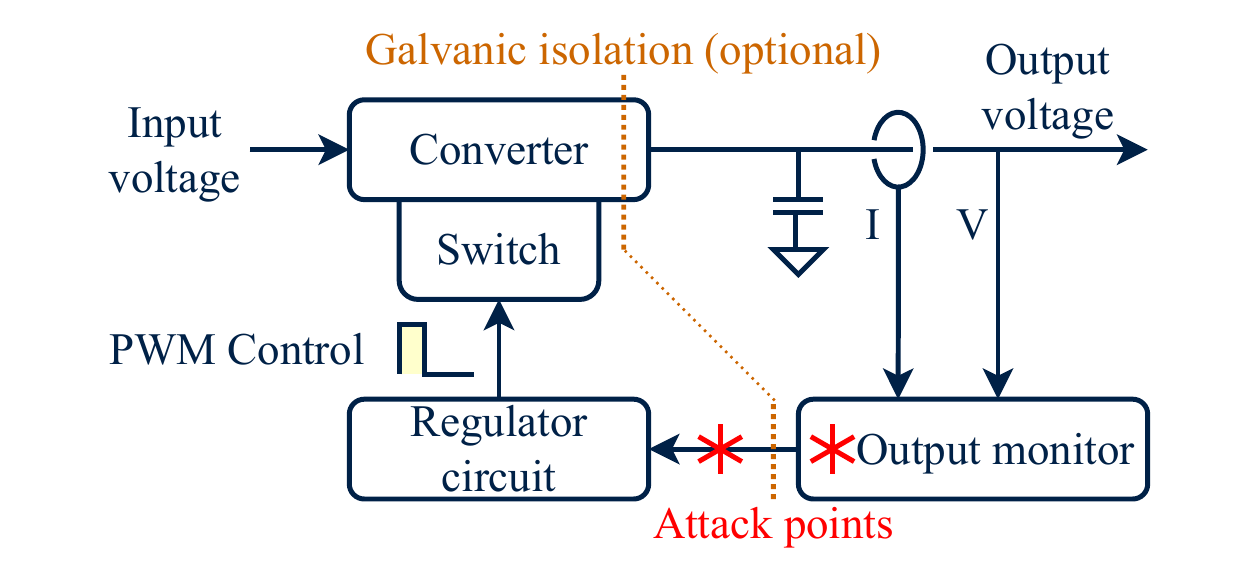}
  \caption{Diagram of a generic switched-mode converter, showing the 4 most important functional components, as well as the most vulnerable part of the circuit where we expect the attacking signal to couple into.}
  \label{fig:smps}
\end{figure}

Power converters take a supply of electricity, whether AC or DC, at a certain voltage and convert it to an output, also AC or DC, at a different voltage.
The two simplest options are a transformer (AC-AC) and a rectifier (AC-DC).
In both cases, the output voltage is a fixed ratio of the input, and can not be adjusted. 
The DC supplied by a simple rectifier also has a high ripple, which is unsuitable for most applications.
A linear regulator (DC-DC) can be used, however they are bulky and very inefficient, thus not suitable for most applications~\cite{horowitz}.

Switched-mode power supplies solve the stability, efficiency and versatility issues, at the cost of added complexity and active components.
They produce a stable output voltage regardless of the input voltage, and the output voltage can easily be adjusted during operation if needed.
They can also easily be modified to supply a constant current, which is important for battery charging, LED~\cite{led-cc} and laser diode applications.
All switched-mode devices consist of four functional components:
one or more inductors and capacitors, which are used as temporary energy storage,
electronically controllable switches,
an output monitoring circuit,
and a regulation system.
A generic schematic of a switched-mode power converter is shown in Figure~\ref{fig:smps}.

The output voltage is determined by the input voltage, duty cycle of the switching, and potentially the load on the output.
Active regulation is thus absolutely necessary, so the output is monitored to derive a feedback signal, and this feedback signal is fed into the regulator, which adjusts the duty cycle of the system accordingly.
By creating a feedback loop which adjusts the output based on the feedback signal, the system is able to maintain the output at a stable value, regardless of changes to the input voltage or load.

It is common for most of the regulation to be integrated into a single IC, and it is not practical to manufacture chips for all conceivable output voltages, nor is it reasonable to make the regulator IC directly monitor large outputs (for systems above \SI{100}{\volt}).
In many cases, such as mains-based power supplies, it is desirable to galvanically isolate the input and output, so the feedback must be conveyed optically.
For these reasons, an output monitor circuit measures the output, and derives a feedback signal, which can then be compared against a fixed or adjustable reference in the regulator.
The feedback signal can be analog or digital, and may be conveyed optically.
The output monitor can be as simple as using a resistive divider or Zener diode to decrease the output voltage to the magnitude of the reference, and conveying this analog signal.
Digital systems will include an ADC and systems which measure the output current will include the necessary current sensing and amplification circuits.

\subsection{Battery Overview}
\label{sec:bg_bat}

\begin{figure}
  \centering
  \includegraphics[width=\linewidth]{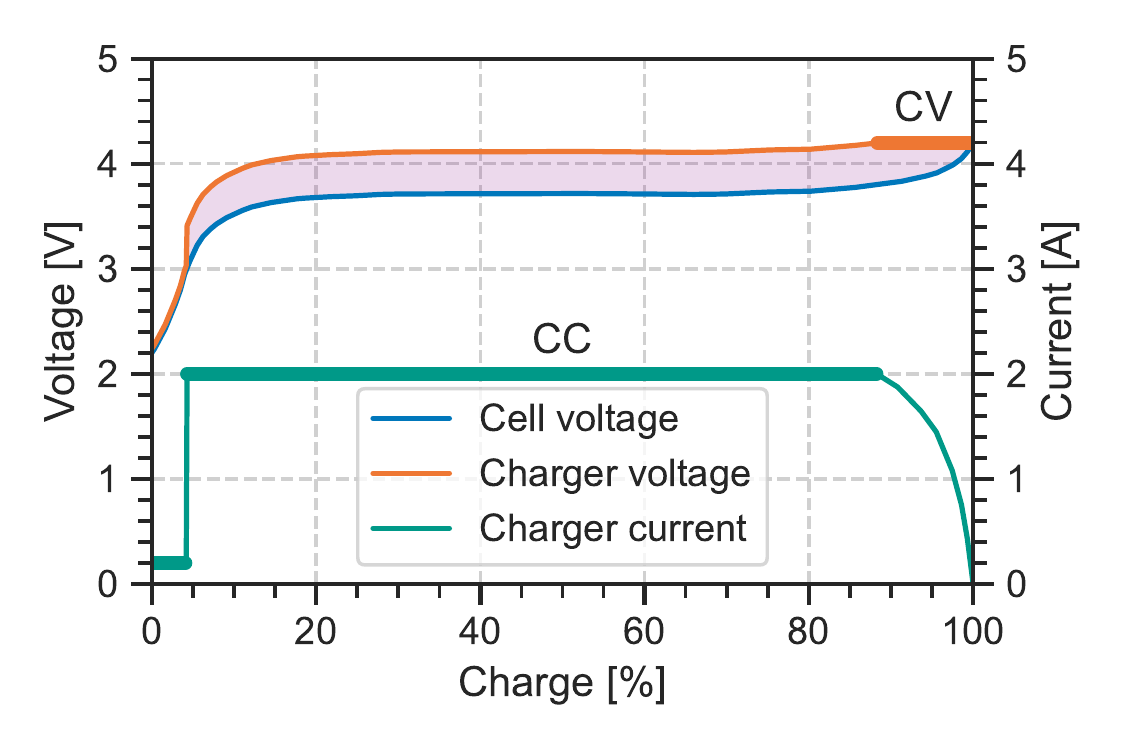}
  \caption{Approximate diagram of the charging phases for a Li-ion battery. The diagram shows the cell's internal voltage, external voltage (including shaded voltage drop across its internal resistance), and the charging current. Bold sections show the limiting current and voltage during the CC and CV phases respectively.
  }
  \label{fig:charge}
\end{figure}

Batteries are a way of storing chemical energy so that it can be easily retrieved in the form of electricity.
Charging or discharging a battery causes a chemical reaction to occur inside the cell, which generates waste heat.
While each battery chemistry has an associated ``nominal'' voltage, the cell voltage changes with the State of Charge (SoC) of the battery.
The exact relationship depends on chemistry, manufacturing technique and can even vary from cell to cell, but for Li-ion it approximately follows the shape shown in Figure~\ref{fig:charge}.

Overcharging the battery occurs when the battery is connected to a voltage greater than the maximum cell voltage, and can lead to an electrical breakdown, short circuit, runaway chemical reaction, and fire.
Discharging a battery past its recommended limits is generally non-violent, but the battery may be damaged permanently if recharged from this state.
Since passing any current through a battery (whether during charge or discharge) requires a chemical reaction that produces waste heat, the battery will get warm during either operation.
Charging or discharging at an excessive current outside of the manufacturer's specification can cause the battery to heat up significantly.
This can cause a so-called thermal runaway and potentially lead to a fire.

We model the battery as a Thevenin equivalent circuit~\cite{thakkar21electrical}: an ideal voltage source $V_\text{cell}$ and an Equivalent Series Resistor $R_\text{ESR}$ in series.
Charging the battery is done by connecting an external voltage $V_\text{charger}$ larger than the internal cell voltage to the battery, this will cause current to flow into the battery.
The magnitude of the current can be calculated using the ESR as:
\begin{equation}
I_\text{charge} = (V_\text{charger}-V_\text{cell}) / R_\text{ESR}
\label{eq:charge_I}
\end{equation}
This relationship is also demonstrated in Figure~\ref{fig:charge}, since the shaded region between the cell and charger voltage is proportional to the current.

\subsection{Charger Overview}
\label{sec:bg_chg}

The charger converts an external power source (e.g., mains AC, 5V USB) to an appropriate voltage for the given battery chemistry, it is thus a form of power converter.
Battery charging is not as simple as connecting the battery to a fixed voltage, since for a discharged battery where the cell voltage is much smaller than the charge voltage, the charge current would be too big, in accordance with Equation~\ref{eq:charge_I}.
Instead, charging begins with a Constant Current (CC) phase~\cite{liion-charge}, where the charger voltage is kept slightly above the cell voltage to ensure that a constant current passes through the battery at all times.
As the cell becomes charged, the cell voltage increases, and so must the external charger voltage, to maintain the constant current.
To prevent overcharging, the charger will switch into the Constant Voltage (CV) phase, where the voltage of the charger is not increased any further, and so as the cell voltage continues to increase, the charge current decreases.
When the charge current is sufficiently low, charging is deemed complete, and all external power can be disconnected.
Chargers may also incorporate a precharge or regeneration phase~\cite{liion-regen}, where if a deeply depleted battery is detected, CC charging is done with a much smaller current, to bring the cell voltage back into the nominal region.
Connecting the full CC charging current to a deeply depleted battery could cause excessive damage, depending on the battery chemistry.
The stages of charging are illustrated in Figure~\ref{fig:charge}.

\section{Attack Theory}
\label{sec:theory}

In this section we provide an overview of the attack, followed by a detailed theoretical evaluation of its components.

\subsection{Attack Description}
\label{sec:theory_desc}
We propose an IEMI sensor attack, where an RF signal is emitted in close proximity to the target.
The wires inside the target act as unintentional antennas, receiving the attack signal.
Since the signal is at a much higher frequency than those used in the target, it has no direct effect, however previous work has shown that non-linear properties in the analog front end of sensor systems (amplifiers, ADCs, comparators) can demodulate this RF signal, converting it into a DC offset \cite{giechaskiel2019framework}.
This effect is strongly frequency dependent, since the receiving ``antenna'' will be resonant at a certain frequency, and the non-linearity also has frequency dependent characteristics.
We use this attack vector to modify the voltage and current measurements needed for the feedback system in switched-mode converters.

Switched-mode converters rely on a feedback system for correct operation.
The device measures its own output, and compares it against the target value.
If for example the output is measured to be lower than it should be, the converter increases the output voltage, thus also increasing the measurement.
Our attack targets the sensors used for this measurement: if an offset is introduced between the real and measured value, and the converter maintains the measurement equal to the target, then the attacker has successfully introduced an offset between the real and target output voltage.
 
\subsection{IEMI Theory}
\label{sec:theory_iemi}

We now consider how non-linear properties of the target can cause the RF signal to demodulate.
The effect observed in real-world devices is the combination of a non-linear element, followed by a low pass filter.
The non-linear effect can be understood by considering a signal $x(t)$ passing through a non-linear element (e.g., diode or transistor) being modified into $f_{NL}(x(t))$.
This new signal can be expressed using a Taylor expansion: $f_{NL}(x(t)) = a + b x(t) + c x(t)^2/2 + \dots$.
The constant and linear term are part of the intended operation of the device, and have no relevance to the attack.
The higher order terms are generally undesirable, and allow for frequency conversion, for example the quadratic term applied to a simple sinusoid will be converted:
$$A \sin(\omega t) \rightarrow c A^2 \sin^2(\omega t) / 2 = c A^2 (1 - \cos(2 \omega t)) / 4$$
By applying a low pass filter, only the constant term is left, hence the non-linearity and low-pass converted the RF signal into a DC offset.
The sign of the constant $c$ is important, as it is the only factor that determines the sign of the attack voltage.
This means that if the attacker wants to shift the measurement in a certain direction, they cannot do so purely by varying power, instead they need to find a frequency and injection point with the appropriate sign.
The power can then be used to adjust the magnitude of the injected signal.

\label{sec:theory_power}
By modeling the receiving wire as an antenna with a very low gain, we can conclude that at a given frequency the received power is $P_R \propto P_T / d^2$ in terms of the transmitter power $P_T$ and distance $d$.
The induced signal will thus have amplitude $A$: $P_R = A^2 / 2Z$ in terms of the impedance $Z$ of the receiver.
Since the induced RF voltage with amplitude $A$ for a second-order effect is converted to a DC signal with $\Delta V_\textrm{attack} \propto A^2$, we conclude similarly to \cite{tu2019trick}:
\begin{equation}
    \Delta V_\textrm{attack} \propto A^2 \propto P_R \propto P_T / d^2
    \label{eq:tx_power}
\end{equation}

\subsection{Regulator Feedback}
\label{sec:theory_feedback}

At its core, any feedback system takes in some measurements of the system, compares them to a reference value, and then issues corrective actions.
The corrective action causes the system to change, which changes the measurements, thus completing the feedback loop.
In power converters, the measurement is related to the output voltage or current, and the corrective action is adjusting the duty cycle of switching, which then influences the output.
As discussed earlier, it is not always possible for the output to be directly measured.
For example, current must first be converted to a voltage, and large voltages must first be decreased before they can be safely connected to an ADC or regulator IC.
These converted values can then be used as measurements in the feedback loop.

This scheme has an additional benefit: it allows the same regulator IC to be used for different output voltages, by only adding some passive external components.
To understand how this works, consider that there is a function converting the output voltage to the measurement voltage $V_{\textrm{meas}} = f_M(V_{\textrm{out}})$.
It is this converted value that is compared against the reference value, and maintained at equality: $V_{\textrm{ref}} = V_{\textrm{meas}}$, regardless of whether the two signals are fed into an analog comparator, or both are digitized and compared in software.
In either case, for this calculation we assume that the attack point is this comparison or digitization, taking $V_{\textrm{meas}} \rightarrow V_{\textrm{meas}} + V_{\textrm{attack}}$, thus in the adversarial case: $V_{\textrm{ref}} = V_{\textrm{meas}} + V_{\textrm{attack}}$.
We can thus express the unattacked and attacked scenarios:
\begin{equation}
    V_{\textrm{out},0} = f_M^{-1}(V_{\textrm{ref}}) \qquad
    V_{\textrm{out}} = f_M^{-1}(V_{\textrm{ref}} - V_{\textrm{attack}})
\end{equation}
We now consider some potential measurement functions, found in the tested devices, and calculate how they are expected to impact the attack.

\subsubsection{Voltage dividers}
A simple method to convert a large output voltage to smaller range is by using a voltage divider, given by $f_M(V) = \beta V$ with $\beta < 1$.
We thus get:
\begin{equation}
    \label{eq:fb_divider}
    V_{\textrm{out},0} = \frac{V_{\textrm{ref}}}{\beta} \qquad
    V_{\textrm{out}} = \frac{V_{\textrm{ref}} - V_{\textrm{attack}}}{\beta} \\
\end{equation}
\begin{equation}
    \label{eq:vout_fix_div}
    \Delta V_{\textrm{out}} = V_{\textrm{out}}-V_{\textrm{out},0} = - \frac{V_{\textrm{attack}}}{\beta} \\
\end{equation}
\begin{equation}
    \label{eq:vout_adj_div}
    \frac{\Delta V_{\textrm{out}}}{V_{\textrm{out},0}} = -\frac{V_{\textrm{attack}}}{V_{\textrm{ref}}}
\end{equation}
By considering adjustable and fixed dividers separately, we can gain further insight.
If the device has an adjustable divider but constant reference, we can use Equation~\ref{eq:vout_adj_div}.
This shows that at a given attack power, the fractional change in the output will always be the same, regardless of the current division, and the current output voltage.
If instead the divider is constant but the system can adjust the reference value, Equation~\ref{eq:vout_fix_div} shows that the absolute offset will be a constant at any set output voltage.

\subsubsection{Zener Diode}
Zener diodes are an electronic component that feature a fixed voltage drop, that can be manufactured to a variety of values.
They are commonly used in AC-DC converters because they can provide a fixed voltage drop with a low impedance, to power the opto-coupler.
They can be written as $f_M(V) = V - V_Z$, thus:
\begin{equation}
    \label{eq:vout_zener}
    \Delta V_{\textrm{out}} = - V_{\textrm{attack}}
\end{equation}
This shows that identically constructed power supplies that only differ in their Zener diode and thus output voltage will be equally impacted by the attack.

\subsection{Current Measurement Systems}
\label{sec:theory_current}

It is standard for all power supplies to have a set output voltage feature, where the voltage will be kept constant regardless of load.
This is unsustainable, as at sufficiently high current (e.g., short-circuit) the device would need to supply infinite power.
Thus, all power supplies also have an inherent current limit, set by fuses or failure, above which it can not operate.
While this is good as a safety feature, in many applications it is desired for power supplies to have a reliable current limiting feature.

In essence, such a device seeks to output the maximum possible voltage, such that the voltage is less than the voltage limit, and the current flowing through the supply is less than the current limit.
If the output is for example short circuited, the voltage will be decreased to near zero, but a current equal to the current limit will still flow.
Such devices are utilized extensively in laser diode and LED drivers~\cite{led-cc}, which are extremely sensitive to changes in their input voltage, and can best be controlled by limiting the current flowing through them.
The CC-CV battery charging scheme (except for the regeneration phase) can also be performed simply using a current and voltage limited power supply, by setting the current limit to the desired CC charge current, and the voltage limit to the CV phase voltage.

In order to implement the current limit, the device must measure the output current.
There are two primary ways to measure current: current shunt and magnetic fields.
In the current shunt approach, the output current $I$ is ran through a small resistor in series with the load, and the voltage difference across the resistor $V_R$ is measured.
The wasted power by the resistor is $P = V_R I$, hence for electric vehicles where $I=\SI{1}{\kilo \ampere}$ is possible, $V_R \approx \SI{1}{\milli \volt}$ is needed to avoid unnecessary losses and heating of the resistor.
In lower power systems, the current will be smaller, but less power is available to be wasted, hence small voltages will always be common.
The second method consists of placing a magnetic field sensor next to the wire, and measuring the current through it's magnetic field.
Both of these methods are good targets for an attacker, since the voltages that are measured are very small, and will be heavily amplified.

Regardless of the details of current measurement, the current is converted into a voltage to serve as the measurement for the feedback system, and we can now use $V_\textrm{meas} = f_M(I)$ in the previous section.
The conclusions derived about how the nature of the measurement and regulation system changes the effect of the attack will still apply.

\section{Evaluation}
\label{sec:eval}

In the previous section we developed theoretical predictions for how different set voltages and loads change the ability of the attacker to impact the operation of the converter.
In this section, we extensively evaluate a large selection of power converters, including DC-DC, AC-DC, constant current and battery chargers.
Where possible, we compare our theoretical predictions with the experiments.

\subsection{Experimental Setup}

\begin{figure}
  \centering
  \includegraphics[width=\linewidth]{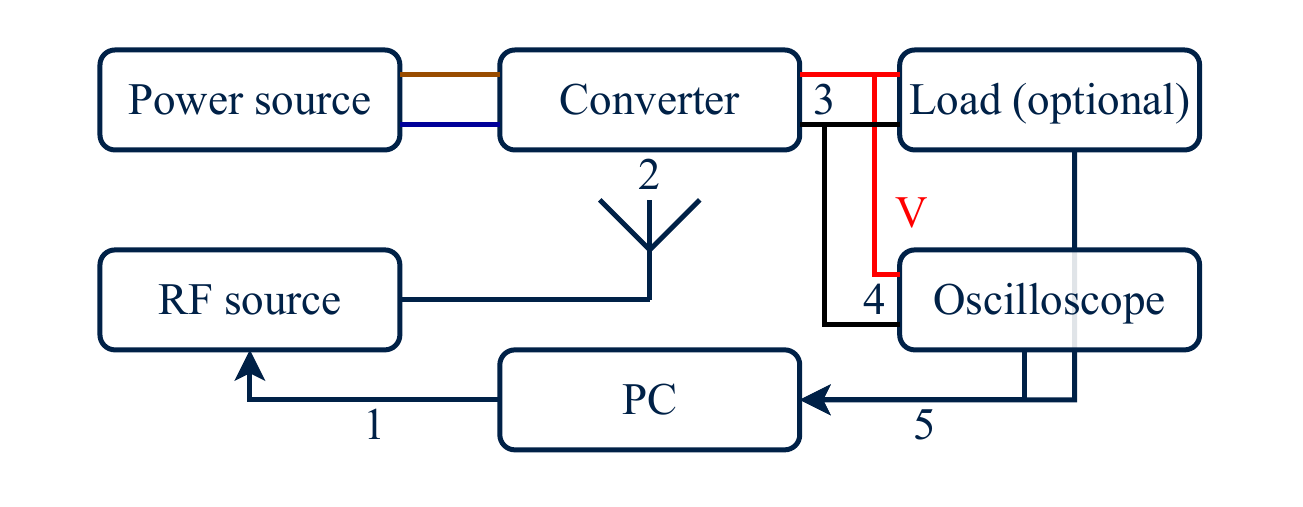}
  \caption{Diagram of the measurement setup. The PC configures the frequency source to the next values (1), the antenna emits the EMI signal at the converter (2), the converter responds to the attack (3), changes are measured by the scope (4) and logged by the PC (5). This repeats for the next frequency.}
  \label{fig:setup}
\end{figure}

We investigate the response of various power converters to a pure sine wave IEMI attack.
To accomplish this, we set up each converter in realistic conditions, with a power source and optionally a load.
An antenna is placed next to the device, to irradiate it with an RF signal.
The frequency of the signal is varied across a wide range, and the output voltage or current is recorded during the frequency sweep.
A vulnerable device will be resonant to some attack frequencies, and this will show up on the sweeps as a peak in the output measurement around a specific frequency.
If possible, multiple subsequent runs are performed, adjusting parameters such as the set output voltage or applied load, without changing the geometry or other parameters of the setup.
A schematic of the attack setup is shown in Figure~\ref{fig:setup}.

\begin{figure}
  \centering
  \includegraphics[width=0.7\linewidth]{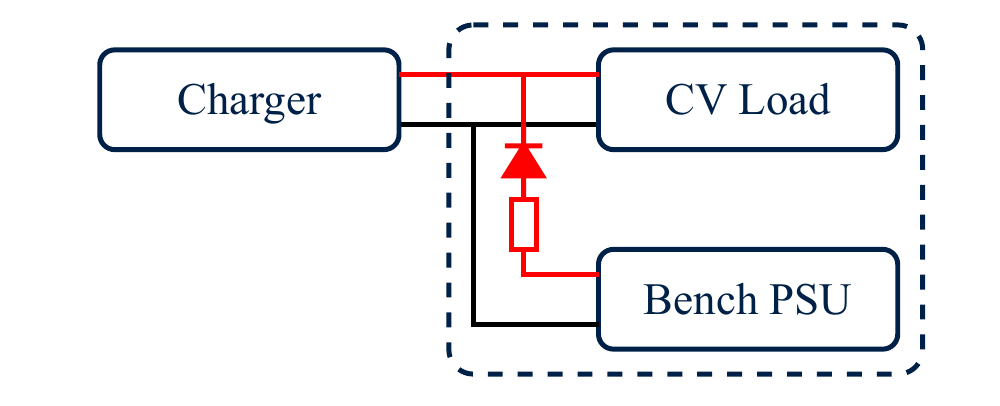}
  \caption{Diagram of the battery simulator. When idle, the DC power supply and load together set the measured voltage. When charging, the load can sink large currents.}
  \label{fig:battery_sim}
\end{figure}

For safety reasons, most experiments were not conducted with the intended load attached, since a successful attack could easily damage the device, or in the case of batteries result in a fire.
Instead, the Rigol DL3021A DC load was connected to the converter, as it can safely accept voltages and currents much larger than those studied in this paper.
The DC load can be programmed to act as a Constant Voltage (CV), Constant Current (CC) and Constant Resistor (CR) load, as appropriate for the given experiment.

\label{sec:battery_sim}
Battery chargers often require the battery to output a voltage when not charging, in order to measure its state of charge, and initiate the process.
A battery simulator was thus created by connecting the load in parallel with a lab bench power supply, as shown in Figure~\ref{fig:battery_sim}.
The DC load was set to CV mode to emulate $V_\textrm{internal}$ of the battery, and a total ESR of \SI{66}{\milli \ohm} is provided by the leads running to it.
This allows the battery simulator to consume large currents while being ``charged'', and the additional lab PSU in parallel allows a small amount of current to be sourced from the simulated battery.

\begin{table*}
    \centering
    \resizebox{\textwidth}{!}{
    \begin{tabular}{lllllllllllll}\toprule
        ID & Input & \multicolumn{3}{c}{Output} & Features & Topology & Main IC & Voltage sense & \multicolumn{4}{c}{Current sense}\\
        \cmidrule(lr){3-5} \cmidrule(lr){9-9} \cmidrule(lr){10-13}
        & & \multicolumn{2}{l}{Voltage} & Current & & & & Feedback & Shunt & Location & Amp & Feedback\\ \midrule
        DC-DC \#1 & \SIrange{5.5}{30}{\volt} & \SIrange{0.5}{30}{\volt} & $\updownarrow$ & < \SI{3}{\ampere} & CC & Single ended SEPIC & FP5139 & Adjustable divider & \SI{25}{\milli \ohm} & Low & LM358 & Adjustable reference\\
        DC-DC \#2 & \SIrange{3}{40}{\volt} & \SIrange{1.5}{35}{\volt} & $\downarrow$ & < \SI{3}{\ampere} & & Buck & LM2596 & Adjustable divider & \multicolumn{4}{l}{N/A} \\
        DC-DC \#3 & \SIrange{5.5}{30}{\volt} & \SIrange{0.5}{30}{\volt} & $\updownarrow$ & < \SI{3}{\ampere} & CC & Single ended SEPIC & FP5139 & Adjustable divider & \SI{25}{\milli \ohm} & Low & LM358 & Adjustable reference\\
        DC-DC \#4 & \SIrange{5.5}{30}{\volt} & \SIrange{1.25}{30}{\volt} & $\updownarrow$ & < \SI{5}{\ampere} & CC & 4-Switch Buck-boost & LTC1625 & Adjustable divider & \SI{7}{\milli \ohm} & Low & LM321 & Adjustable reference\\
        DC-DC \#5 & \SIrange{8}{36}{\volt} & \SIrange{1.25}{32}{\volt} & $\downarrow$ & < \SI{5}{\ampere} & CC & Buck & XL4015E1 & Adjustable divider & \SI{50}{\milli \ohm} & Low & LM358 & Adjustable reference\\
        DC-DC \#6 & \SIrange{10}{40}{\volt} & \SIrange{1.2}{36}{\volt} & $\downarrow$ & < \SI{40}{\ampere} & CC & 2-Switch Buck & LM5116 & Adjustable divider & \SI{4}{\milli \ohm} & Low & LM321 & Adjustable reference\\
        \bottomrule
    \end{tabular}
    }
    \caption{Test of 6 different DC-DC converters. $\uparrow$ indicates that the output voltage must be higher than input, $\downarrow$ means lower, and $\updownarrow$ either. For Constant Current (CC) capable devices, the method of current sensing, and current sense amplifier are provided.}
    \label{tab:dcdc}
\end{table*}

The voltage and current can be monitored directly on the load, but a Rigol DS2302A Oscilloscope is connected in parallel, to provide high quality measurements.
Using an oscilloscope in this environment is advantageous, as they are designed to accept RF input signals, and are thus unlikely to be influenced by attacks.
In the paper always the most appropriate and accurate data source is used to present the plots, however the auxiliary data sources are checked to ensure that the effect is real and not a result of attacks on the instruments themselves.
The RF signal is supplied by a Rhode \& Schwartz SMC100A Signal generator, which can output \SI{19}{\dBm} (\SI{80}{\milli \watt}) up to \SI{3.2}{\giga \hertz}.
A PC is connected to all the instruments, changing their settings and collecting the data.

Unless otherwise stated, the signal is supplied directly by the signal generator and radiated by a Vert-900 monopole antenna.
We used the maximum \SI{19}{\dBm} power from the source, sweeping the frequency range \SI{50}{\mega \hertz} to \SI{3}{\giga \hertz}.
The lowest frequencies were skipped because they were far out of range for the antenna, whereas the upper limit was added because high frequencies are strongly attenuated by parasitic properties of everyday materials, unless they have been specifically designed for microwave operation.

\subsection{DC-DC Converters}
\label{sec:eval_dcdc}

\begin{figure}
    \centering
    \includegraphics[width=\linewidth]{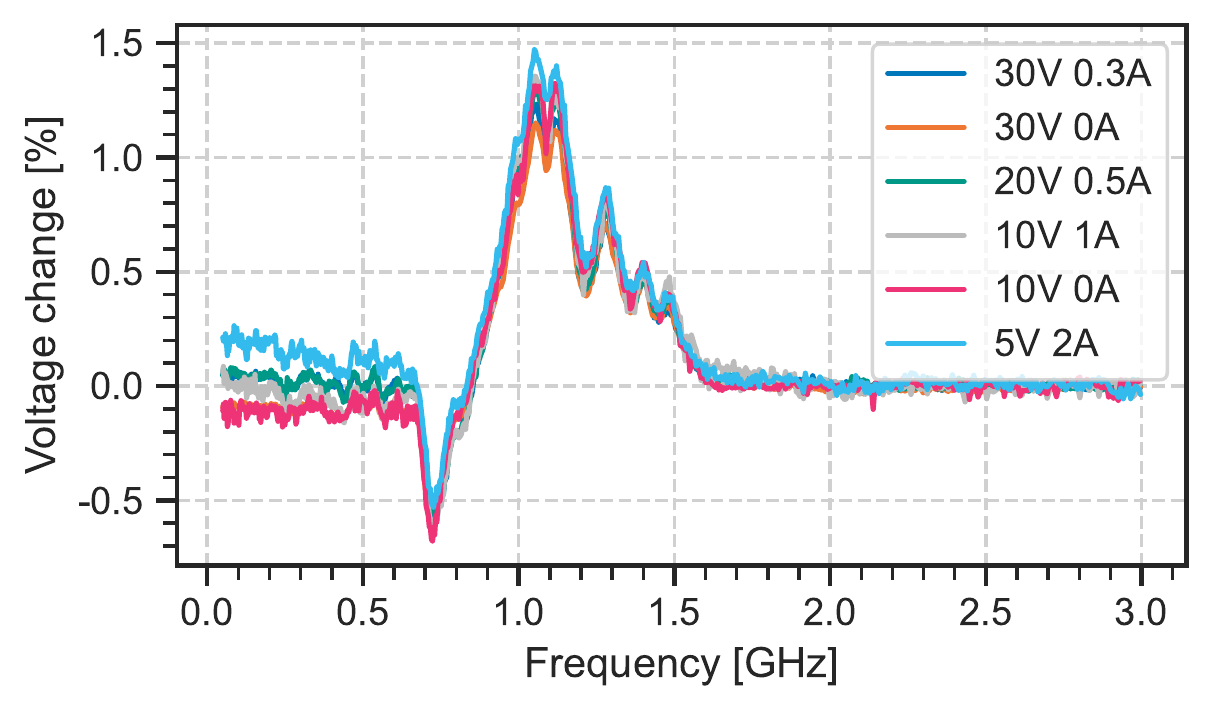}
    \caption{Relative change in the measured output voltage of DC-DC \#1. Different lines represent different set output voltages and applied loads. Input is \SI{12}{\volt}, some measurements are in boost, some are in buck operation. Confirming the theory, voltage and load do not affect the fractional change.}
    \label{fig:dcdc_V_id1}
    \label{fig:dcdc_V}
\end{figure}

DC-DC converters occur in end user products, where a single DC input (e.g., 5 or \SI{12}{\volt}) is converted to appropriate levels for all internal components, such as \SI{3.3}{\volt} logic ICs, lower voltages for modern processors, or various voltages for analog circuits.
Such converters are often based on reputable industry standard regulator chips, following the recommended application circuit in their datasheet.

To evaluate the security of these converter regulator chips and designs, we tested various prototyping non-isolated DC-DC converter boards, available from commercial websites.
The intended audience for these devices is electronics prototyping, and the circuits on them also follow closely the recommended application circuit that would be found integrated into the circuitry of commercial devices.

A list of the converters tested and their relevant features can be found in Table~\ref{tab:dcdc}.
The key difference between these versatile prototyping devices and their deployed counterpart is that the prototyping devices use an adjustable resistor to change the output voltage, whereas the end product will use fixed resistors.
However, this does not influence the outcome of experiments.
The use of an adjustable resistor allows us to test the same circuit operating at different set output voltages, and compare it against the theory developed.
Due to the adjustable resistor-based architecture, we expect the operation to be described by Equation~\ref{eq:vout_adj_div}.

\subsubsection{Method}
The devices tested are sold as raw PCBs, allowing us to place the transmitting antenna in close proximity to the regulator IC.
For each device, it was connected to an input DC power source, an output voltage was selected, and a Constant Resistor (CR) load or no load was applied to the output.
A CR load was chosen as it allowed the output voltage to vary in response to the attack unlike a CV load, and is much simpler than a CC load.
An antenna was placed next to the converter, in close proximity to the regulator IC, and a frequency sweep was carried out with \SI{19}{\dBm} power, in the \SI{50}{\mega \hertz} to \SI{3}{\giga \hertz} range.
For each converter, multiple output voltages and loads were tested, while the input voltage and geometry are kept constant.
These experiments thus examine the attack impact as a function of the set output voltage, and eliminate variety due to any other parameters.

\subsubsection{Results}

Figure~\ref{fig:dcdc_V} shows the fractional change in output voltage for device \#1 at various set outputs and loads.
Plots for other devices can be found in the Appendix.
The plot shows that the fractional change is independent of the set output voltage and applied load, validating the theoretical expectation derived in Equation~\ref{eq:vout_adj_div}.
We thus prove that the primary injection point is into the feedback reference pin, instead of the output voltage line.
This is also shown for converters \#3 and \#6, while \#5 shows that the effect is constant regardless of output voltage but changes with load.
Converter \#2 has a different impact at different voltages and loads, however the effect is always visible at the same frequency.
Finally, converter \#4 had no measurable change within our tested ranges.

Our results demonstrate that many typical DC-DC converters are vulnerable to IEMI attacks, with all vulnerable devices showing a stable impact, and constant resonant frequency regardless of output settings.
For multiple devices we also show that the theoretical expectation holds, giving a constant fractional change regardless of setting or load.

The order of magnitude of the attack impact was around \SI{1}{\percent}, with \SI{19}{\dBm} power, however more power would allow an attacker to produce a larger difference.
For many applications \SI{10}{\percent} deviations would be within tolerances, however, DC-DC converters are also used to power CPU and GPU chips, which are very sensitive to small changes in voltage \cite{dcdc-vrm}.
We thus conclude that with higher attack power, or by targeting sensitive devices, the attack can achieve significant impact.

\subsection{AC-DC Converters}
\label{sec:eval_acdc}

\begin{figure}
    \centering
    \includegraphics[width=\linewidth]{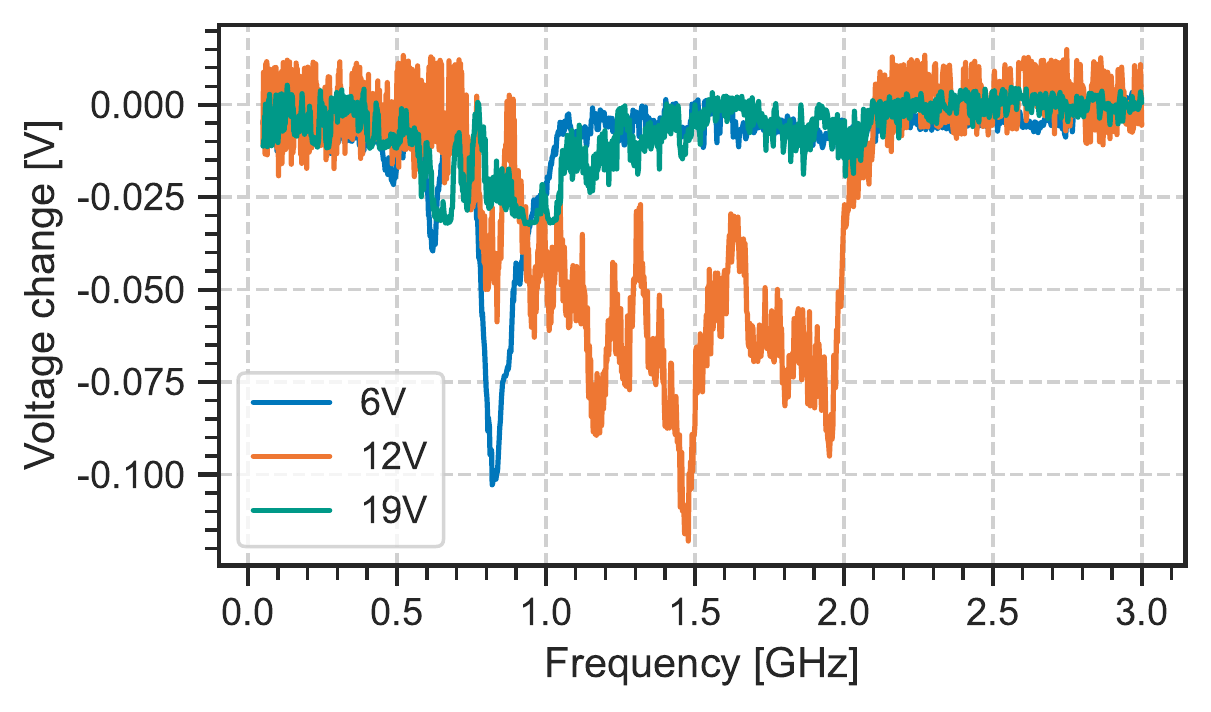}
    \caption{Attack against three different AC-DC wall plugs. All devices show a measurable response to the attack with only \SI{19}{\dBm} power.}
    \label{fig:acdc}
\end{figure}

Most devices receive their power from AC mains, through AC-DC converters.
The architecture of these devices is substantially different from DC-DC converters, for example, due to their full galvanic isolation, and optical feedback system.
To examine the security of these devices we tested three power bricks sold with commercial devices.
Fixed voltage AC-DC converters usually feature a Zener diode architecture, thus we would expect a constant voltage offset, for multiple converters that differ only in their output voltage.
However, the converters tested had different manufacturers, shapes and sizes, thus no comparison between them is possible.

\subsubsection{Method}
The antenna was first moved around to find the best location, where the attack signal can easily couple onto the most sensitive component.
The devices were not connected to a load during these experiments, and no modifications were made to the housing of the converter, in order to ensure a realistic attack.
The housing introduces a separation between the internal components and the antenna, and is thus expected to lower attack success with the same \SI{19}{\dBm} power.

\subsubsection{Results}
Figure~\ref{fig:acdc} shows that all tested devices were impacted by the attack, but unlike the DC-DC converters their output voltage decreased, likely due to a systemically different architecture:
the voltage at the feedback pin of DC-DC regulator ICs increases with increasing output voltage, whereas in AC-DC converters commonly using an N-type opto-coupler feedback system, the feedback voltage decreases when the output voltage increases.
Assuming that the sign of $\Delta V_\textrm{attack}$ induced in the feedback input is the same in both systems, it is expected that the impact on the output will have an opposite sign due to the opposite nature of feedback.

\subsection{Constant Current Supply}
\label{sec:eval_cc}

\begin{figure}
    \centering
\includegraphics[width=\columnwidth]{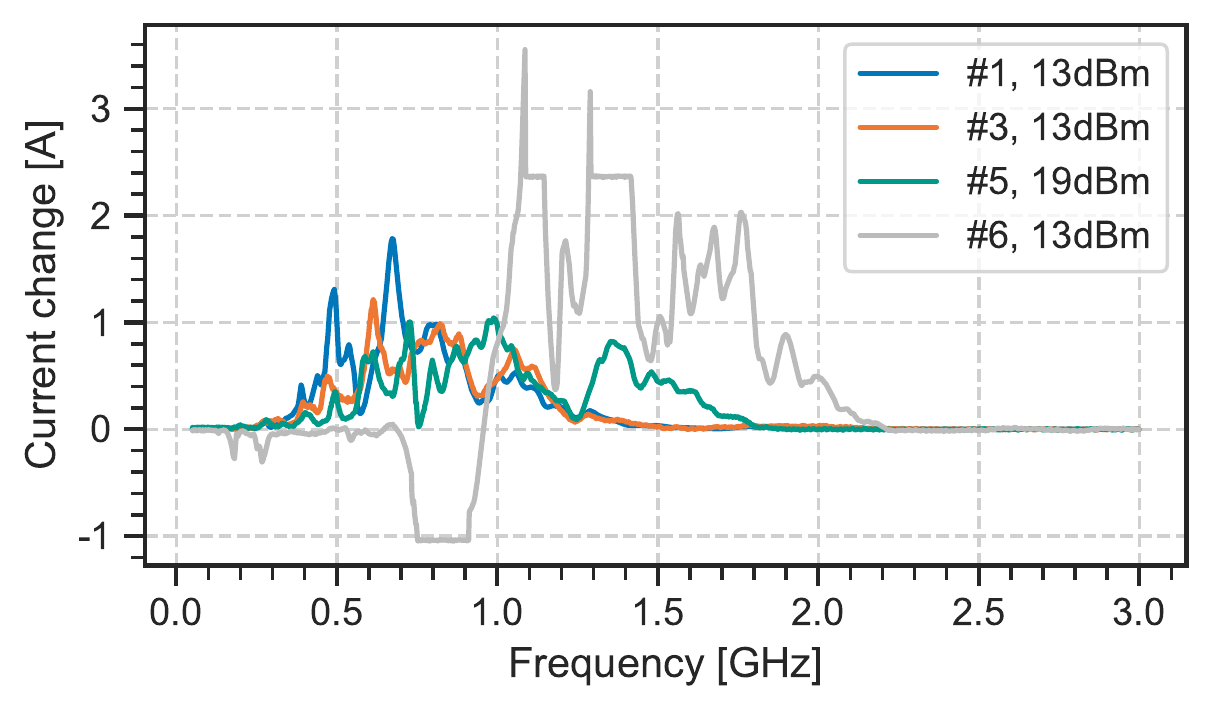}
    \caption{The DC-DC converters set to \SI{1}{\ampere} constant current mode, with a \SI{7}{\volt} CV load applied. The devices are all seen to be highly vulnerable to the attack. We had to decrease the attacker power for three of them, to get a meaningful plot.}
    \label{fig:dcdc-cc}
\end{figure}

All but one of the devices in Table~\ref{tab:dcdc} featured a constant current (CC) output capability.
This means that the device will not exceed the current limit on its output, and is an essential part of the CC-CV battery charging scheme.
As explained in \ref{sec:theory_current}, it is expected that the current sensor is significantly more sensitive than the voltage sensor, and we thus expect the impact of the attack to be higher.

\subsubsection{Method}
The devices were set to constant current mode, by increasing the output voltage limit as high as possible, and decreasing the current limit to \SI{1}{\ampere}.
A CV load at \SI{7}{\volt} was applied below the voltage limit of the supply, to ensure that the output is current limited and not voltage limited.
The input was chosen to be \SI{12}{\volt}, as this is a common input voltage to many systems.
The \SI{1}{\ampere} limit was chosen to be high, but within the safe operating region of all devices, and the \SI{7}{\volt} voltage was chosen because it is characteristic for two discharged Li-ion cells in series.
The stable operating point of such a system is that the current flowing into the load is the current limit set on the supply, and the voltage is the CV value set on the load.
The use of a CV load is representative of batteries and LEDs, two common systems which are driven by a constant current, since the voltage drop across them changes very little as a function of current.
A monopole antenna was placed next to the device, transmitting a pure sine wave with \SI{19}{\dBm} or lower power in the \SIrange{0.05}{3}{\giga \hertz} range.

\subsubsection{Results}
Figure~\ref{fig:dcdc-cc} shows the effect of attacking the devices while in CC mode.
For three devices (\#1, \#3, \#6), the attack power had to be reduced to measure a useful frequency response curve, since the response was so powerful that the increased current risked damage to the converter, or overloaded the power supply powering the converter.
The effects of this overload can be seen as the flat plateaus for device \#6, even at \SI{13}{dBm}.
The strong response particularly for device \#6 can be explained by the very small shunt resistor used.
This is advantageous when the device is rated for a high maximum current, to decrease wasted power, however greatly increases the capability of an attacker.
Device \#4 claimed to feature a CC mode, however it was too unstable to use, regardless of the attack.
The experiments showed a much larger change in current compared to the change in voltage experienced by the same devices during a CV attack, easily achieving multiples of the set current, and reaching the limit of the devices.
Due to the importance of CC supplies in battery and LED systems, this attack demonstrates a significant vulnerability.

We also examined how the attack works at various loads.
Since the devices used an adjustable reference scheme for current control, we expect the impact to be governed by Equation~\ref{eq:vout_fix_div}.
We tested device \#6, since it had the lowest ripple, and thus most stable operation.
The results on Figure~\ref{fig:dcdc-cc-loads} clearly show that the absolute change in current is independent on the set CC limit or the applied CV load, in accordance with the theory.

\subsection{Automotive Current Sensors}

\begin{table}
    \centering
    \resizebox{\linewidth}{!}{
    \begin{tabular}{lllll}\toprule
        Part number & Range & Method & Output & Sensitivity\\
        \midrule
        TLI4971-A025T5-U & $\pm\SI{25}{\ampere}$ & Hall effect & Analog & \SI[per-mode=symbol]{20.83}{\milli \ampere \per \milli \volt}\\
        TMCS1100A4 & $\pm\SI{5.75}{\ampere}$ & Hall effect & Analog & \SI[per-mode=symbol]{2.5}{\milli \ampere \per \milli \volt}\\
        INA253A3 & $\pm\SI{6}{\ampere}$ & Shunt & Analog & \SI[per-mode=symbol]{2.5}{\milli \ampere \per \milli \volt}\\
        INA260 & $\pm\SI{15}{\ampere}$ & Shunt, ADC & Digital & \SI[per-mode=symbol]{1.25}{\milli \ampere \per LSB}
    \end{tabular}
    }
    \caption{Summary of automotive grade current sensor tests.}
    \label{tab:current_sensors}
\end{table}

\begin{figure}
    \centering
    \includegraphics[width=\columnwidth]{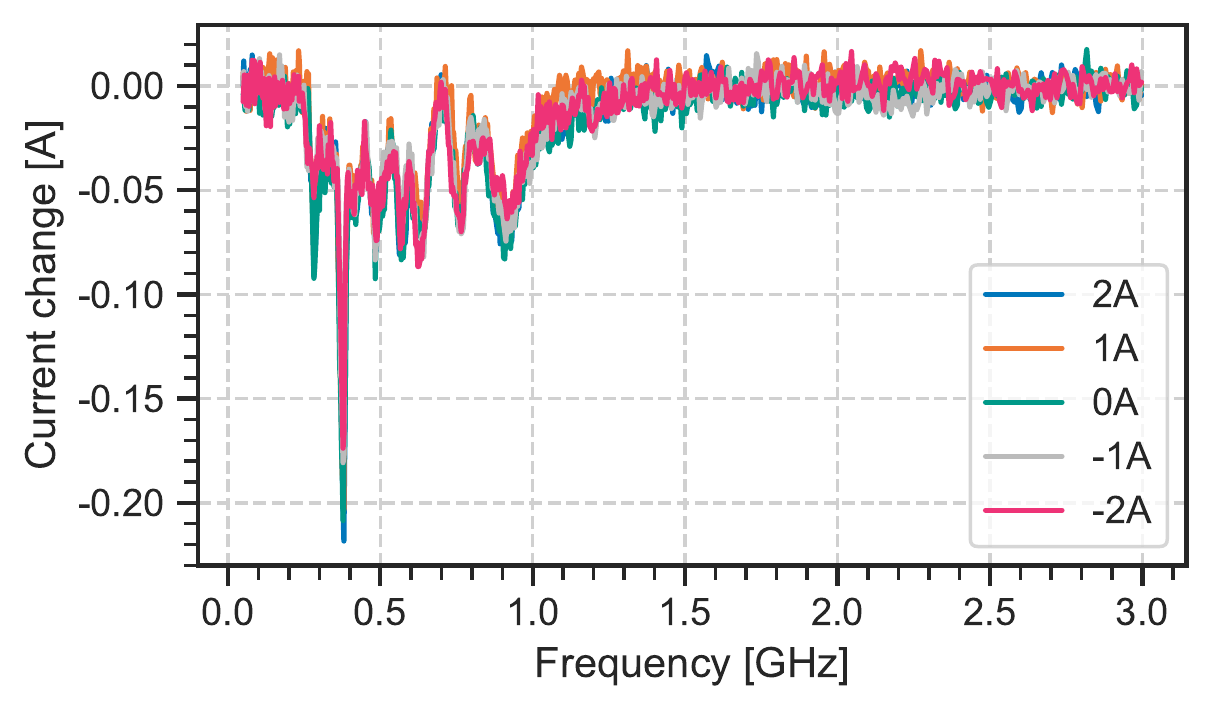}
    \caption{Attack against the TLI4971 magnetic current sensor, at different actual current values. The effect of the attack is a constant offset, regardless of actual current.}
    \label{fig:cs1}
\end{figure}

The previously examined CC drivers all relied on a shunt-based current sensor, with a separate discrete amplifier.
In many industrial applications, small, tightly integrated, galvanically isolated or digital sensors will be used.
We thus tested four different current sensors from reputable vendors, marketed for applications such as EV Supply Equipment, photovoltaic inverters, or other forms of power infrastructure.
The selected devices include magnetic and shunt-based technologies as well as analog and digital outputs. Their key properties are summarized in Table~\ref{tab:current_sensors}.

\subsubsection{Method}
A DC lab bench supply was set to \SI{3}{\volt}, and a CC load at various current values was applied.
The CC load ensures that the current being drawn from the supply is always at the constant, set value, thus creating a stable current value for the measurements.
The \SI{3}{\volt} limit is chosen to minimize the wasted power, while being high enough to overcome any resistance in the wires between the supply and load.
Since the measurement focuses on current sensors, the voltage used to generate the currents has no influence on the attacks.
The sensor is connected between the two devices, on the low-side, and an antenna is placed near it.
To measure negative currents, the two inputs of the current sensors are swapped.
In previous experiments, the attack caused the real voltage or current in the system to change, with plots showing real changes recorded by non-attacked sensors.
However during this experiments, the real current does not change, and any measured change is only due to attacks on the sensor used for measurement.

\subsubsection{Results}
Figure~\ref{fig:cs1} shows the effect of the attack against the TLI4971 sensor, other plots can be found in the Appendix.
In this case, the measured value is decreased, and by a constant amount regardless of the sensor current.
The TMCS100 sensor shows a current dependent positive effect always at the same frequency.
The INA260 shows both positive and negative effects at different frequencies, with some of the frequencies showing constant offsets, and others being current dependent.
Finally, the INA253 showed no change due to the attack.

Notably, in previous experiments, an increase in the output voltage or current was the higher impact result, since this increased voltage can cause damage to the downstream devices.
This is not the case for sensor attacks, where lowering the measurement is the significant result.
This is because the lowered measurement will cause the feedback system to increase the output, in order to bring the measurement back to the desired value.
Thus, lowering the sensor measurements in a feedback system is necessary to increase the output voltage, thus executing the more damaging attack.

By demonstrating a successful attack against automotive grade sensors, we demonstrate that power converters used in these industries are vulnerable to attacks, since a spoofed sensor measurement causes the feedback loop to deviate from the desired output voltage or current values.

\subsection{Battery Chargers - CC Phase}
\label{sec:eval_charger}

\begin{figure}
    \centering
    \includegraphics[width=\linewidth]{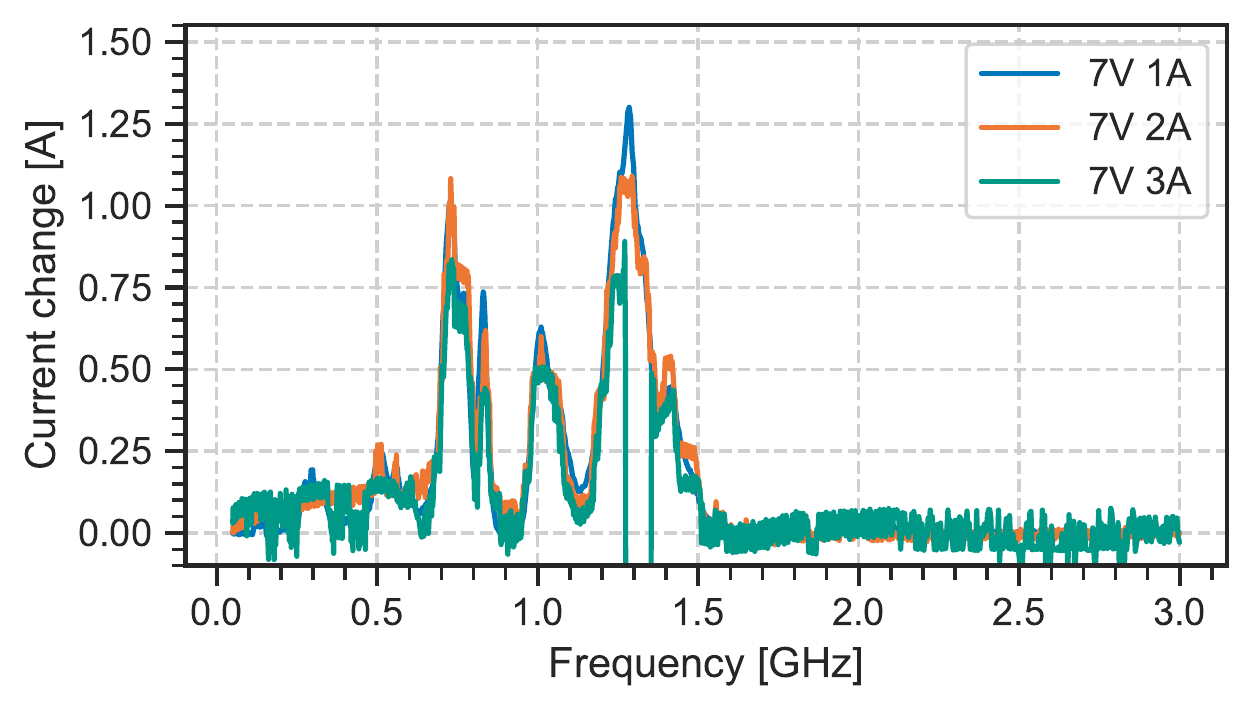}
    \caption{Attacks against the SkyRC e4 battery charger for the three different current settings. The results show a constant offset regardless of set charge current.}
    \label{fig:chg_current}
\end{figure}

The number of devices being powered by batteries is rapidly increasing.
Battery chargers are CC-CV supplies, however they often feature additional features such as protection systems, or shutoff when charging is deemed complete.
Due to fundamental similarities between normal CC-CV supplies and battery chargers, we expect that the attacks demonstrated against them will also work against battery chargers.
Because of their additional protection features and complexity, we experimentally examine chargers to see if they are equally vulnerable to the proposed attacks.

Charging a battery with a larger current than normal can cause excessive heating and damage, while reaching a larger charge voltage can also cause permanent overcharge damage.
For safety reasons, the experiments were initially conducted without a real battery, instead using the battery simulator described in Section~\ref{sec:battery_sim}.
Using a simulator also allows for increased accuracy and repeatability, since a simulated battery can stay at a constant state of charge for the entire experiment.
Besides providing accurate measurements, the use of a battery simulator ensured that any malfunction would not cause a safety risk or damage.

We picked two off-the-shelf battery chargers commonly used for charging drone batteries.
More specifically, we tested the SkyRC e4 and C240 Duo battery chargers.
While the SkyRC e4 charger only allowed the charging of two or more cells in series with 1, 2 or \SI{3}{\ampere}, the C240 was able to charge one or more cells with up to \SI{10}{\ampere} in \SI{0.1}{\ampere} steps.
In both cases, multiple charger settings were tested, to determine how they effect the results.

\subsubsection{Method}
The charger was connected to the battery simulator, set to emulate an appropriate cell voltage for the batteries it would normally be charging: we assumed Lithium ion batteries, and set a cell voltage of \SI{3.5}{\volt} per cell, representing a mostly discharged battery that needs to be charged in the CC phase.
The SkyRC charger required a minimum of two cells in series, and access to the intermediate cell voltage for balancing.
In this case, a simple resistive divider was added across the terminals of the battery simulator to create the intermediate cell voltages.
Due to the divider, all of the simulated cells have the same voltage across them, thus the balancing feature is not used, and the simulation should not impact the result.
A good location for the antenna was determined manually, before frequency sweeps were performed.
In line with previous experiments the output power of the attack signal was set to \SI{19}{\dBm}.
When investigating the impact of charger settings on the attack success, the antenna was stably fixed to the device to ensure that the geometry of the attack is not a variable.

\subsubsection{Results}
Figure~\ref{fig:chg_current} shows the results of the frequency sweep against the SkyRC e4 charger for different current settings.
The plot shows that for this charger, the attack always achieves the same constant offset, regardless of the selected current setting.
The results also indicate that the charging current through the battery can be significantly increased by \SI{1}{\ampere}, by an attacker transmitting only \SI{19}{\dBm} (\SI{80}{\milli \watt}).
A large increase in charge current will cause excessive heating in the battery, decreasing it's lifespan, or potentially causing more significant damage.
The charger was also seen to re-boot when attacked during the \SI{3}{\ampere} experiment, indicating that the increased power exceeded its internal design limits.

The plot for the C240 Duo can be found in the Appendix.
It also shows an increase of \SI{0.1}{\ampere} using only \SI{19}{\dBm}, hence this charger is also vulnerable to the attack.

\subsection{Battery Chargers - CV Phase}
\label{sec:eval_charger_cv}

\begin{figure}
    \centering
    \includegraphics[width=\linewidth]{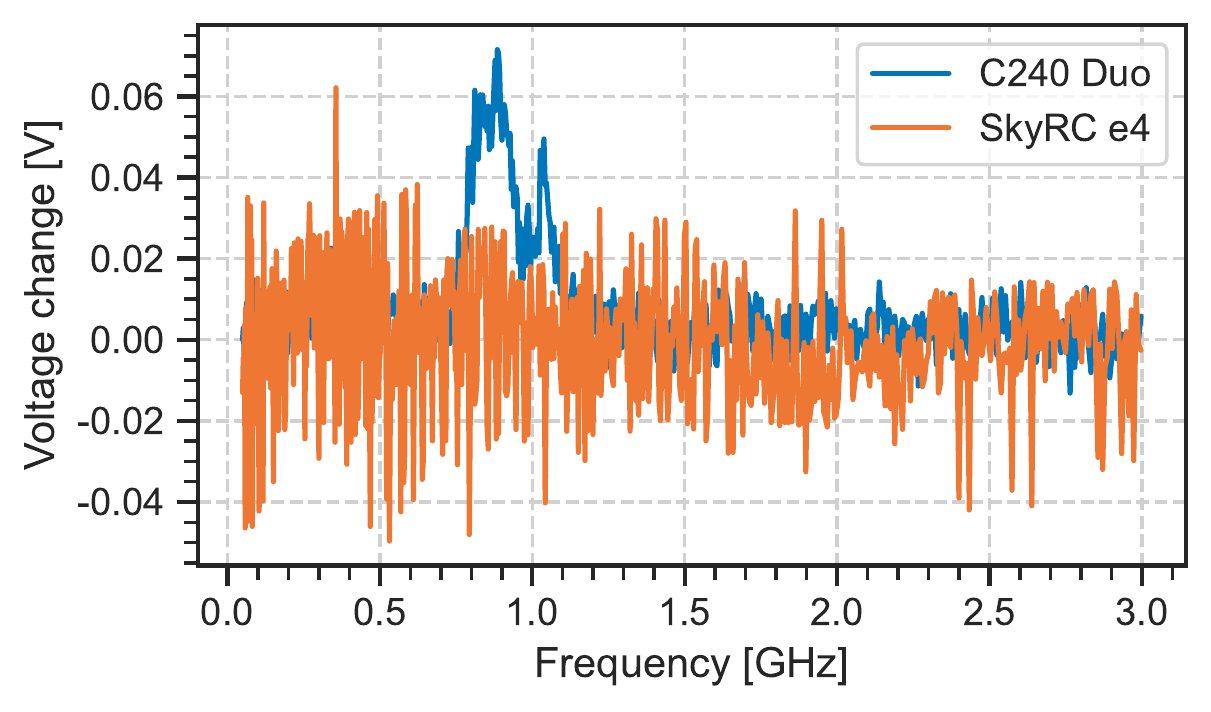}
    \caption{Attacks against the two chargers in the CV phase, only the C240 is shown to be vulnerable in this range.}
    \label{fig:chg_voltage}
\end{figure}

Next we evaluate chargers in the CV phase of charging.
We expect that voltage sensors are less vulnerable to the attack, and we thus expect a less significant response to the attack.

\subsubsection{Method}
We seek to examine how the attack effects the chargers perception of the cell voltage, as this will effect the voltage reached by the charger at the end of the CV phase.
In a real charging scenario, the attack would begin, thus causing the charger to perceive a low cell voltage.
The charger will thus remain in the CC phase of charging, even when the cell voltage exceeds the set value.
Eventually, the cell voltage become sufficiently high that one of the following outcomes occurs:
an auxiliary protection mechanism shuts off charging,
the battery fails due to an overcharge,
the charger finally perceives the voltage to reach the CV limit, and charging stops.
We could replicate this process by increasing the simulated cell voltage of our battery simulator for every frequency, until charging stops, and then manually re-starting the charger.
However, in the interest of time, we modified the setup as follows:
We used the previously described battery simulator, and connected a high power \SI{3.7}{\ohm} resistor in series with it.
Charging was started, and due to the high ESR, the charger immediately entered the CV phase.
We then decreased the internal cell voltage, thus increasing the current, while ensuring that the charging is still CV limited.
This scheme allows the charger voltage to increase or decrease, with only a small impact on the current.
For the C240 Duo charger, the steady state was \SI{4.2}{\volt} \SI{0.3}{\ampere} at the charger, and \SI{3}{\volt} on the load.
For the SkyRC charger, it was \SI{8.4}{\volt} \SI{1.5}{\ampere} at the charger, and also \SI{3}{\volt} on the load.

\subsubsection{Results}
The frequency response of the two devices can be seen in Figure~\ref{fig:chg_voltage}.
Our results show that the SkyRC charger is not measurably vulnerable at \SI{19}{\dBm}, however the C240 shows a clear resonance peak.
If the attack were scaled to \SI{39}{\dBm} (\SI{8}{\watt}), increasing in power 100 times, the C240 attack would correspond to a \SI{6}{\volt} difference.
Overcharging by this amount is a significant attack, while the power level is still within the budget of hobbyists.

\subsection{Attacker Power Budget}
\label{sec:eval_pow}

\begin{figure}
    \centering
    \includegraphics[width=\columnwidth]{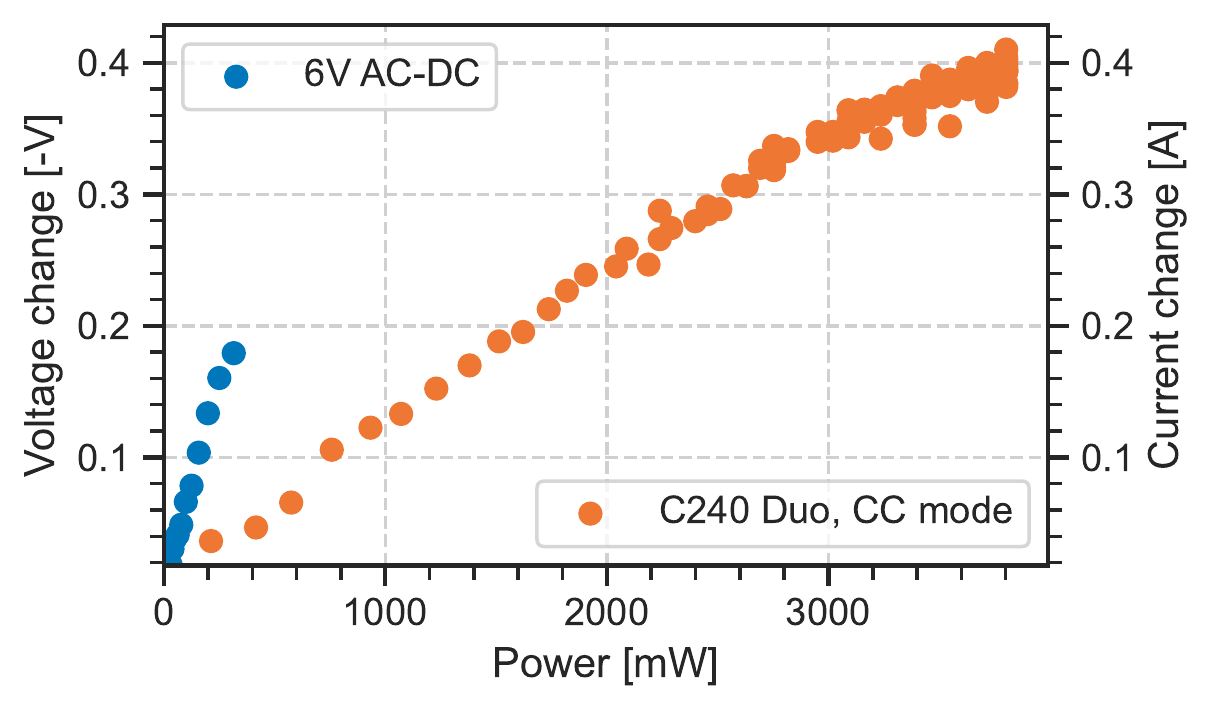}
    \caption{The impact of attacker power against two devices, both showing the expected linear trend.}
    \label{fig:powers}
\end{figure}

The attacker will seek to minimize their budget, and thus the required transmission hardware.
In particular we envision that an attacker may have access to a wide-frequency low-power source (software-defined radio or lab RF signal source) to find the suitable resonance frequency of the target, and will then build or purchase a frequency-specific power amplifier.
This may be significantly cheaper than purchasing a wideband power amp in the first place.
The attacker also wants to minimize the power they are emitting, to the minimum required for the desired effect, both for stealth and cost reasons.
Finally, the attacker may seek fine-grained control over their attack value, hoping to precisely tune its impact.

We thus investigate how the attacker can extrapolate their low-power measurements to determine the power level required for a given impact.
According to the theory (Section~\ref{sec:theory_power}), in every case $\Delta V_\textrm{out} \propto \Delta V_\textrm{attack}$, and based on Equation~\ref{eq:tx_power}, $\Delta V_\textrm{attack} \propto P_T$.
It is thus expected that the change in output voltage should be linear with the attacker transmit power, assuming everything else is constant.

\subsubsection{Method}
We investigate two of the previous devices at their resonant frequency, the 6V AC-DC power converter at \SI{820}{\mega \hertz} and the C240 Duo in CC mode at \SI{1.71}{\giga \hertz}.
An additional amplifier was connected between the signal generator and the antenna, allowing a larger power range to be tested.
Instead of sweeping the frequency, the frequency was fixed at a known resonant value for each device, and the output power of the RF source was swept.
The power sweeps were performed, once with a power meter connected to the output of the amplifier to compensate for amplifier gain characteristics, and once with the antenna connected measuring the impact of the attack on the target device.

\subsubsection{Results}
Figure~\ref{fig:powers} shows the change in output voltage compared to the non-attacked case, as a function of output power.
The plot validates our theoretical prediction, that the attack impact is always linear as a function of RF output power, assuming everything else is kept constant.
The attacker can use this knowledge to determine the power they require to execute an attack, or even to precisely vary the impact of their attack during the process, such as to evade countermeasures based on sudden changes.

\subsection{Ranged Experiments}
\label{sec:eval_range}

Theory predicts that the RF power density at the receiver is the only relevant metric to determine the success of the attack.
This can be achieved by a nearby low powered attacker, or a distant high power attacker.
We examine attacks against a battery charger with only equipment available to motivated hobbyists.

\subsubsection{Method}
We used the SkyRC e4 charger, as it demonstrated a larger attack response.
The charger was connected to the battery simulator, charging a simulated battery consisting of two cells totaling \SI{7}{\volt}, with \SI{1}{\ampere}.
We used the Mini-Circuits ZHL-10W-202S+ amplifier, outputting \SI{38.3}{\dBm} (\SI{6.7}{\watt}), connected to an RFSpace TSA600 directional antenna.
We tested the attack at the known resonant frequency of \SI{1290}{\mega \hertz}.

\subsubsection{Results}
In steady state, the charger set to \SI{1}{\ampere} was actually outputting \SI{1.1}{\ampere}.
The change in current depends on the direction of the antenna, and notably the maximum impact did not happen when the antenna was pointed directly at the charger, presumably due to multipath propagation.
We tested our attack from a variety of distances:
from \SI{5}{\meter} the current increased to \SI{1.18}{\ampere}, from \SI{2}{\meter} it increased to \SI{1.5}{\ampere}, and from \SI{1}{\meter} it reached \SI{3}{\ampere}, in approximate agreement with the inverse square law.
We conclude that with this hardware the attack can have a significant impact up to two meters away, which demonstrates a significant real-world threat.

\section{Battery Charging Attacks}

\label{sec:eval_charger_boom}
There may be important differences between the battery simulator used in previous experiments, and a real battery.
We also want to assess whether the increase in current and voltage caused by the attack is sufficient to permanently damage a battery.
We thus tested a standard ``18650'' lithium cell and a small pouch lithium battery, to evaluate their performance during a charging attack.

\subsection{Ethical and Safety Considerations}

As described in Section~\ref{sec:bg_bat}, overcharging a battery can lead to safety-critical situations, such as fires and explosions.
Safety was therefore taken very seriously and we worked with a national government agency to minimize potential safety risks during the experiments.
The experimental setup was placed in a closed blasting chamber within a building specifically designed for explosives testing.
Throughout the experiments, we were physically separated from the chamber and located in a control room from where we monitored the experiments via live telemetry.
Conducting the experiments in this environment also ensured that contamination of the surrounding environment due to leaking chemicals is limited.

Examining the true state of the internal battery chemistry is difficult and internal protection mechanisms might have been destroyed during the experiments.
Because of the chemical reactions that can occur even after charging has stopped, we kept the batteries in the chamber and monitored the temperature for a further 12 hours.
To avoid contamination after the experiments, we stored the batteries in special containers and disposed of them at a recycling facility.

\vspace{-0.4em}
\subsection{Experiment Setup}
Unlike previous experiments, the RF signal was provided by a USRP N210 with an SBX daughterboard, and amplified by a Mini-Circuits ZHL-10W-202S+, capable of delivering up to \SI{40}{\dBm} of power.
The oscilloscope used during these experiments was the Rigol MSO5104, and we chose the C240 Duo charger, due to its versatile output range, and ability to charge single cells.
The C240 is a realistic target due to its implementation of various protection mechanisms, such as overcharge, overvoltage, and overcurrent protection, and its capability to monitor the temperature and SoC.
The transmitting antenna was placed below the charger, and two \SI{2.5}{\meter} long, \SI{5}{\square \milli \meter} cross-section wires from the charger connected it to the battery, with a resistance of \SI{45}{\milli \ohm} each, which had a negligible impact.
The voltage was monitored next to the charger, while the current was monitored by using the long wire as a shunt, both voltages being measured by the oscilloscope.
We attached K-type thermocouples to the batteries, to monitor their temperature during the attack.

Two batteries were tested, the INR\_18650\_30\_E \SI{3000}{{\milli \ampere \hour}} 18650 cell, rated for \SI{4.2}{\volt} max voltage and \SI{900}{\milli \ampere} max charge current, and the Renata ICP402035 \SI{195}{{\milli \ampere \hour}} pouch battery rated for the same voltage and \SI{195}{\milli \ampere} current.

\vspace{-0.4em}
\subsection{Overcurrent Experiment}
The current limit of a battery is chosen by the manufacturer for a reason, and exceeding it will reduce cell life or potentially pose a safety risk.
Therefore, our initial experiments targeted the current sensor in the charger.
The battery was discharged until the cell voltage was around \SI{3.5}{\volt}, and thus charging commenced in the CC phase.
By attacking the current sensor at a resonant frequency of \SI{1.65}{\giga \hertz} and power of about \SI{35}{\dBm}, the charge current was increased, however, due to the internal resistance of the cells, the voltage measured by the charger reached the voltage limit, causing the transition into the CV phase.
Once in the CV phase, further increasing the attack on current sensors does not increase the charging current or voltage.
Charging continued with a current larger than the current limit of the battery, and \SI{4.2}{\volt} across the cell.

The 18650 cell was set to charge with \SI{0.9}{\ampere}, and increased to about \SIrange{1.5}{2}{\ampere}, while the pouch battery set to charging with \SI{0.1}{\ampere} increased to about \SI{0.4}{\ampere}.
Our results demonstrate that under an IEMI attack, the charger can deliver more current than configured into a real battery.
However, due to the internal resistance of the cell the current did not increase significantly above the set value.
While we did not observe any direct negative effects due to the overcurrent charging, we demonstrated that the attacker can exceed manufacturer limits, and thus have the potential for damage.

\subsection{Overvoltage Experiment}
We then conducted experiments by attacking the voltage sensor inside the charger.
Initially, the batteries charged in the CC phase as expected, since real and perceived voltage were below the \SI{4.2}{\volt} threshold.
As the battery voltage approached this value, we initiated the attack by emitting the attack signal at the resonant frequency of the voltage sensor, \SI{855}{\mega \hertz}.
This lowered the voltage as perceived by the charger.
Eventually, the battery voltage reached the threshold, but the charger perceived it to be lower, and carried on the CC charging phase.
As the cell voltage increased above the limit, we kept increasing the attack power up to approximately \SI{33}{\dBm}, and thus increasing the difference between the real and measured value up to \SI{1.3}{\volt}.
As a result, the charger continued to charge the battery to \SI{5.5}{\volt}, while believing that it reached \SI{4.2}{\volt}.
While we only achieved a maximum voltage of \SI{5.5}{\volt}, this limitation was not related to available transmission power, but instead was caused by the attack triggering other unintended effects in the charger, such as pressing buttons or triggering the overtemperature protection despite no thermal sensor being connected.
Such side effects could be overcome by careful study of the frequencies, to determine which frequencies are the least likely to have unintended effects.

As a result of the overcharging, the 18650 battery cell suffered complete and permanent failure.
The cell voltage of the battery climbed till it reached \SI{5.3}{\volt}, and then began decreasing while charging continued.
The cell temperature started to increase, reaching about \SI{80}{\celsius} after approximately two hours, when its voltage suddenly collapsed to zero, rendering it unusable.
This battery was thus successfully destroyed by our attack.

The pouch battery reached a cell voltage of \SI{5.5}{\volt}, which fell back to \SI{4.5}{\volt} when charging was disconnected.
The battery bulged slightly, but completed a capacity measurement test with approximately its nominal capacity.
\section{Countermeasures}
\label{sec:countermeasure}

We present a variety of countermeasures that increase the barrier for a successful attack significantly and discuss their practicality.

\subsection{Redundant Voltage Protection System}

The power input of an appliance, or the battery module in a device can be equipped with a protection circuit.
This circuit can be constructed to monitor the voltage and current, and cut off the device from the outside world using a switch such as a solid state FET or a mechanical relay in case any issue is detected.
Besides monitoring for over voltage and over current, this system can also monitor for a decrease in input voltage (brown-out), internal temperature of the device, reverse voltage (such as from user error) and protect against electrostatic discharge.
Besides preventing against attacks, this protection system also protects against legitimate failure of the power supply, as well as user error connecting an incorrect power supply.
Such systems often partially exist in real-world devices, particularly brown-out detection, as a decrease in input voltage occurs naturally as the device is disconnected from power.

Battery modules inside mobile phones feature a highly advanced circuit board, which is responsible for monitoring State of Charge, and battery health, and also comes with protection elements.
EV battery packs contain an advanced BMS monitoring all cell voltages and the pack current, and features high current contactors to physically disconnect the battery in case of any failure.

As these protection systems themselves rely on accurate measurements of the voltage and current, they are vulnerable to identical sensor injection attacks. However, attack complexity increases as the adversary now needs to attack two subsystems in parallel.

\subsection{Shielding}

An established countermeasure against IEMI attacks is to apply shielding around the circuit.
This will attenuate the RF signals, and thus greatly increase the attackers power budget.

Shielding is done by adding a conductive (metal) layer around the device.
The effectiveness of the protection depends on the amount of shielding added: even a light mesh can attenuate some signals, particularly at lower frequency, but a thick copper housing will achieve superior performance.
In many cases, however, it is desirable for the final product to be cheap, small and light, and thus limiting the amount of shielding that can be practically added.
Additionally, input and output wires require a hole in the shielding. The attack signal can also couple onto these wires, and be carried into the device as conducted interference.

\subsection{Filters}

Given the relatively low frequency at which switching and feedback occurs (\SI{100}{\kilo \hertz}), low pass filters can be used to block the high frequency RF signals that the attacker uses, right before the input of the amplifier/regulator ICs.
Such a filter is already incorporated in many of the existing devices, with small capacitors often added to the feedback line.
Their main purpose is not to protect against an RF attack, but instead to reduce noise on the input, and to tune the feedback loop stability.
As shown by our results, the capacitors already present are not sufficient to prevent the attack.

A major issue with this approach is that low pass filters become ineffective well above their rated cutoff, due to parasitic properties of the components used~\cite{parasitic-filter}.
A filter effective in a large frequency range needs higher quality RF components and design, and thus might be prohibitively costly.
The small section of wire connecting the filter to the input of the sensitive analog component, or the internals of the component itself, will still be vulnerable.

The second issue is that negative feedback systems must maintain loop stability: by nature, a feedback system measures the difference between the output and the desired value, and takes corrective steps to reduce the difference.
If there is a large delay between taking the corrective step and its measurable effects, the system may overshoot the target value, and corrective action must now be taken in the other direction.
In an improperly designed feedback system, this can lead to oscillations.
Adding a capacitor to the feedback line introduces a time delay to the feedback signal, potentially de-stabilizing the circuit, further complicating the design process.

\subsection{Attack Detection}

As laid out, fully preventing IEMI attacks is difficult.
Therefore, academic literature has proposed various approaches to instead detect attacks.
The authors of \cite{zhang2020detection, zhang2022detection} have investigated ways in which systems can detect if they are under IEMI attack.
This can, for example, be accomplished by transmitting and measuring two identical signals and looking at the difference \cite{zhang2022detection}, or by randomly connecting/disconnecting the legitimate input and checking if the disconnected measurement produces an expected ``null'' value \cite{zhang2020detection}.
Since the vulnerable logic components of a feedback system are often inside a tightly integrated chip, these countermeasures require the engineering of new semiconductor devices.
In more software-based systems, where the feedback control is performed by a microcontroller, the designer should be able to integrate such schemes with few added external components.

\subsection{Thermal Fuses}

In many cases, the main damage is caused by an increased voltage leading to an increased current. 
This increased current is then dissipated as heat in a component, causing permanent structural damage.
By fitting devices with an appropriate fuse, the damage caused by overcurrent can be mitigated.
In contrast to sensor-based current monitoring systems, a simple heat-based fuse is not vulnerable.

Similarly, Polymeric Positive Temperature Coefficient (PPTC) \cite{pptc} devices can be used as re-settable fuses.
When the current reaches a trigger point, these resistors heat up and increase their resistance dramatically.
When the current is greatly reduced to below a hold current, the fuse cools down and resets.
By placing the fuse in thermal contact with the device it is protecting, heating of the device can also cause the fuse to trip. Many battery cells include a PPTC to protect against large currents, however, since the rated discharge current is often much higher than the rated charge current, the PPTC is designed to trip above the discharge limit.

\section{Conclusion}
\label{sec:conclusion}

Our paper studied for the first time the real-world effects of IEMI attacks on power converters.
Our findings show that the output sensing, active feedback-based nature of switched-mode power converters allows IEMI attacks to be successfully executed on a wide range of devices.
We demonstrated a measurable attack against a variety of devices with only \SI{19}{\dBm}, well within the budget of a motivated hobbyist, and show that we can achieve a significant impact from up to \SI{2}{\meter} with \SI{38.3}{\dBm}, opening up the possibility for ranged attacks.
We evaluated our attack in realistic settings, causing permanent damage to a Li-Ion battery with about \SI{33}{\dBm} (\SI{2}{\watt}) of RF power.
We thus conclude that IEMI attacks against power converters pose a significant risk, with potentially dangerous real-world impact even at the budget of motivated hobbyists.
Better funded adversaries using high power amplifiers and directional antennae can scale the attack to larger distances in a straightforward manner.
We discuss a variety of countermeasures, highlighting prior academic work in attack detection and sensor authentication systems and physical-layer defenses.

\section*{Acknowledgments}
We would like to thank Orlando Summermatter and the testing group of armasuisse Science + Technology for their support during the experiments.
Marcell and Sebastian were supported by the Engineering and Physical Sciences Research Council (EPSRC).

\bibliographystyle{ACM-Reference-Format}
\bibliography{bib}

\appendix
\section{Additional plots}
\label{app:plots}

In this section, we present the frequency response curves for the devices that were not included in the main body of the text due to space constraints.

\subsection{DC-DC converters}

\label{app:plots_dcdc}
\begin{figure*}
    \vspace{2em}
    \centering
    \begin{subfigure}{0.49\linewidth}
        \centering
        \includegraphics[width=\textwidth]{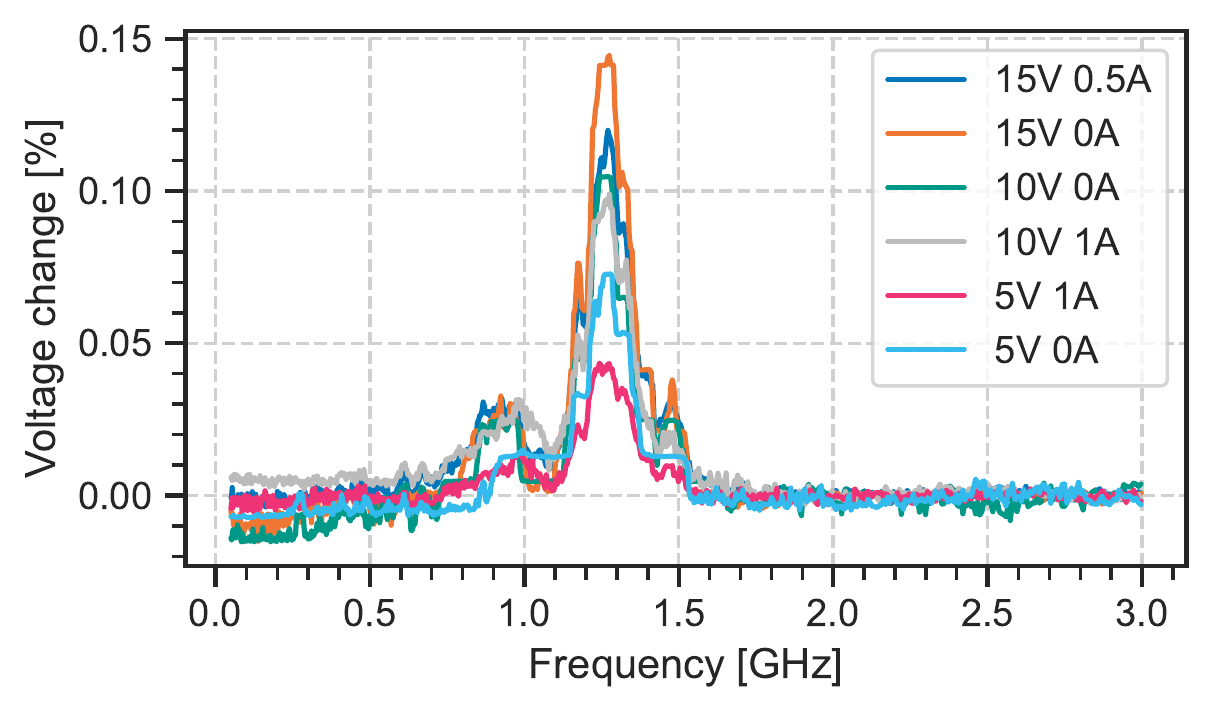}
        \caption{Device DC-DC \#2, \SI{20}{\volt} input.}
        \label{fig:dcdc_V_id2}
    \end{subfigure}
    \begin{subfigure}{0.49\linewidth}
        \centering
        \includegraphics[width=\textwidth]{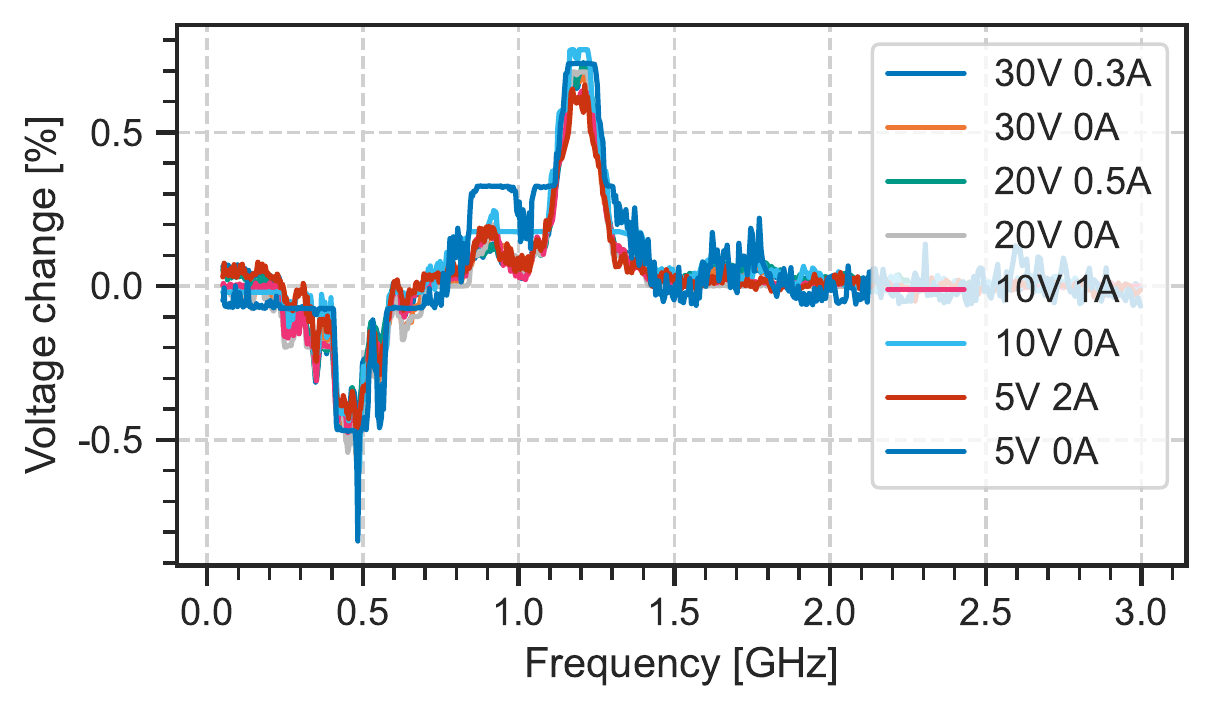}
        \caption{Device DC-DC \#3, \SI{12}{\volt} input.}
        \label{fig:dcdc_V_id3}
    \end{subfigure}\\
    \begin{subfigure}{0.49\linewidth}
        \centering
        \includegraphics[width=\textwidth]{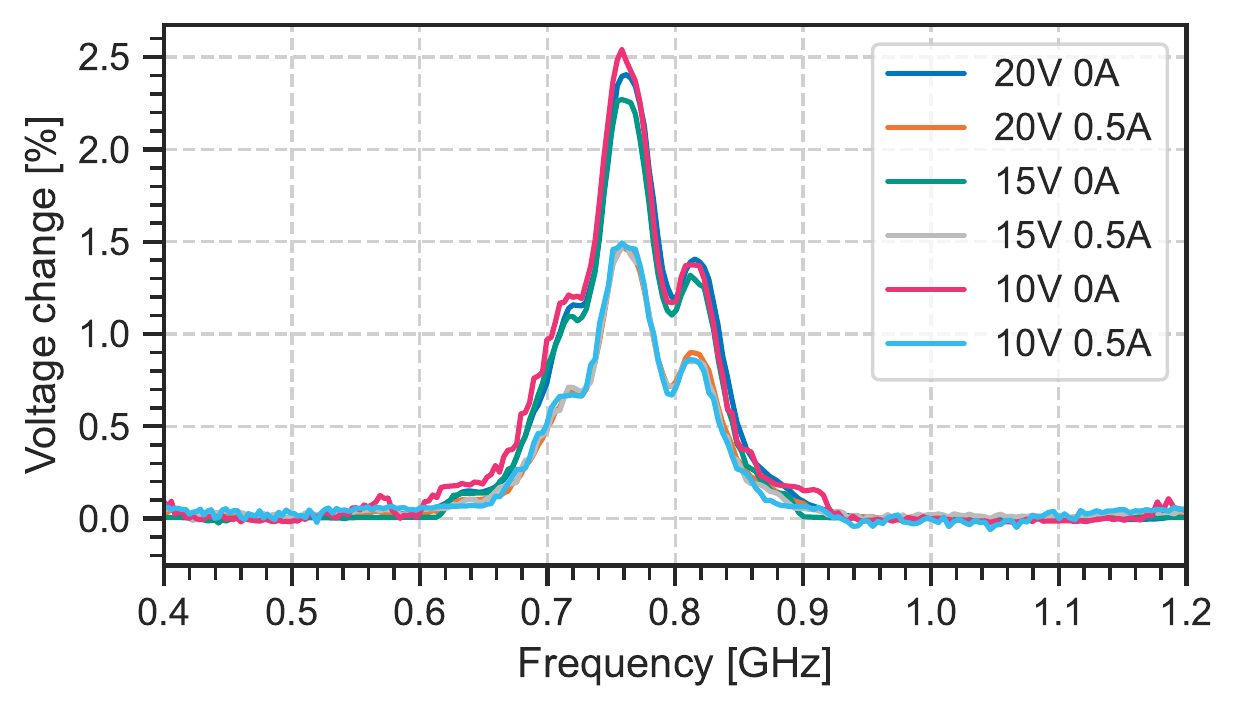}
        \caption{Device DC-DC \#5, \SI{25}{\volt} input.}
        \label{fig:dcdc_V_id5}
    \end{subfigure}
    \begin{subfigure}{0.49\linewidth}
        \centering
        \includegraphics[width=\textwidth]{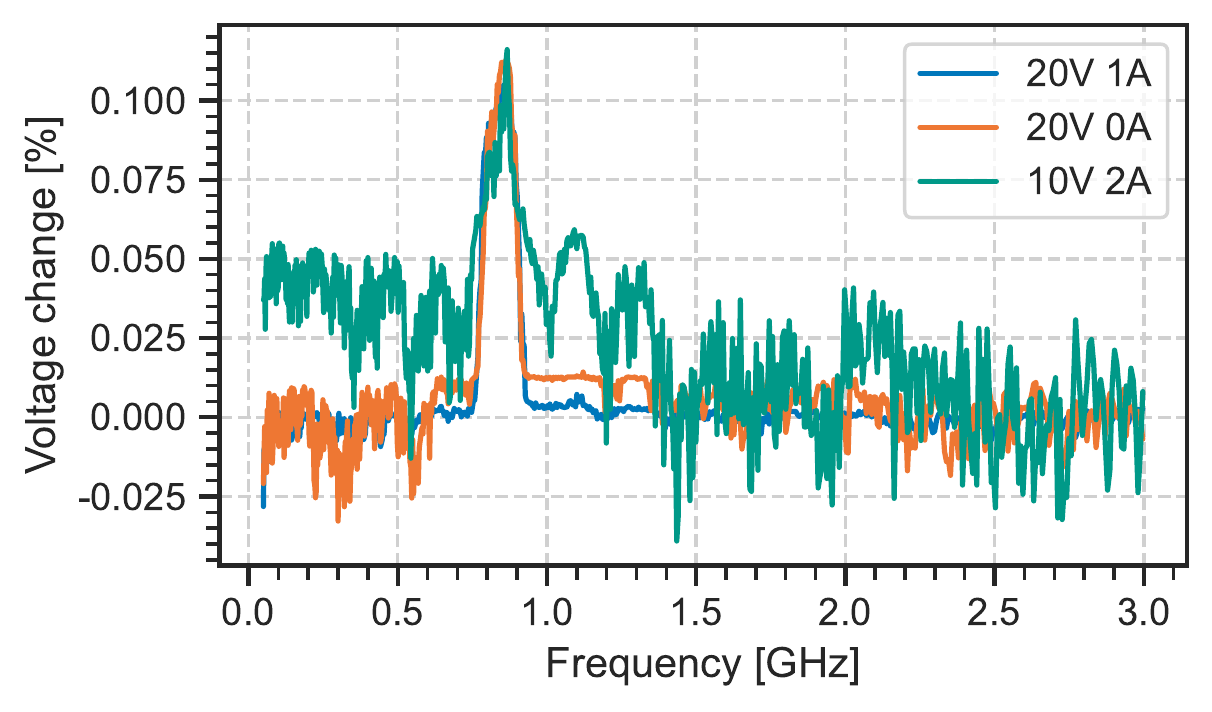}
        \caption{Device DC-DC \#6, \SI{25}{\volt} input.}
        \label{fig:dcdc_V_id6}
    \end{subfigure}
    \caption{Relative change in the measured output voltage on the remaining prototyping boards. Different lines represent different set output voltages and applied loads.}
    \label{fig:dcdc_V_extra}
\end{figure*}

In total we purchased 6 DC-DC converters, however only the results of \#1 are plotted in the main text of the paper.
Each converter had an adjustable output voltage, so we tested multiple output values as well as applied various loads.
As explained in Section~\ref{sec:eval_dcdc}, device \#4 showed no measurable change, and thus is not plotted.
Figure~\ref{fig:dcdc_V_extra} shows the remaining four plots.
Devices \#3 and \#6 reinforce the theory that the attack impact is independent of both the set voltage and applied load.
Device \#5 shows an attack impact that does not depend on the voltage, but depends on the current, and \#2 depends on both.

\subsection{Current sensors}

\label{app:plots_cs}
\begin{figure*}
    \vspace{2em}
    \centering
    \begin{subfigure}{0.49\linewidth}
        \centering
        \includegraphics[width=\textwidth]{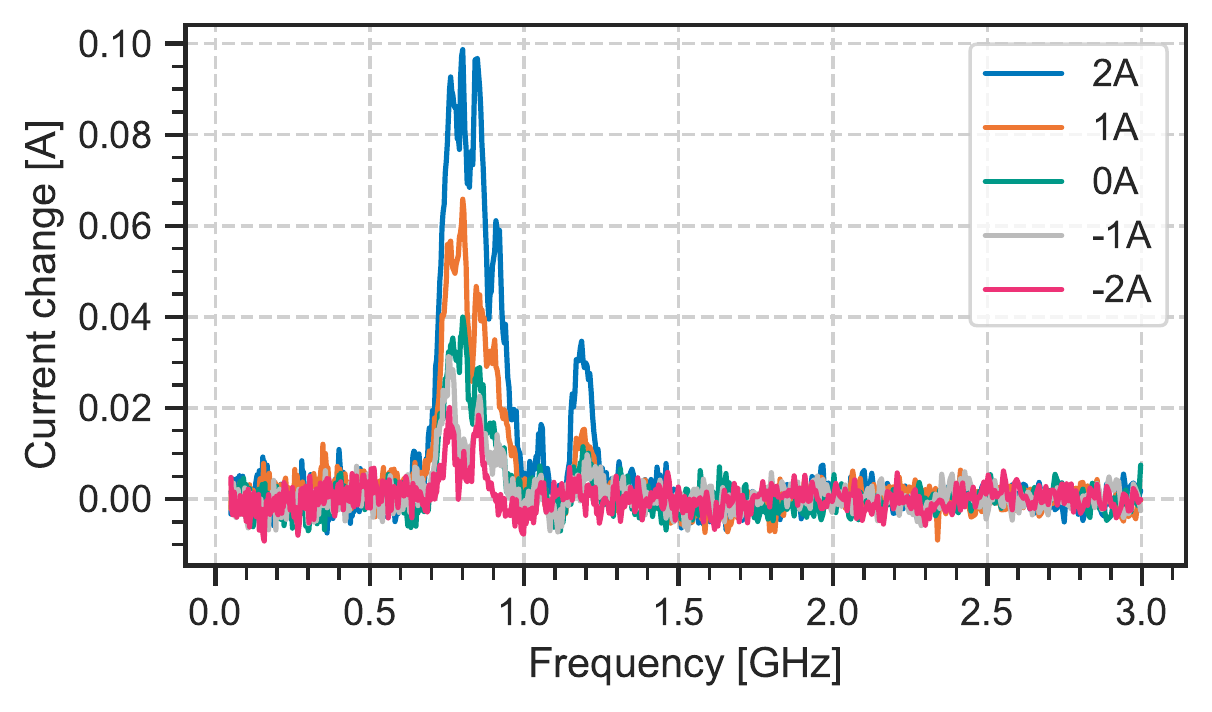}
        \caption{Current sensor TMCS1100 at various real currents.}
        \label{fig:cs_id2}
    \end{subfigure}
    \begin{subfigure}{0.49\linewidth}
        \centering
        \includegraphics[width=\textwidth]{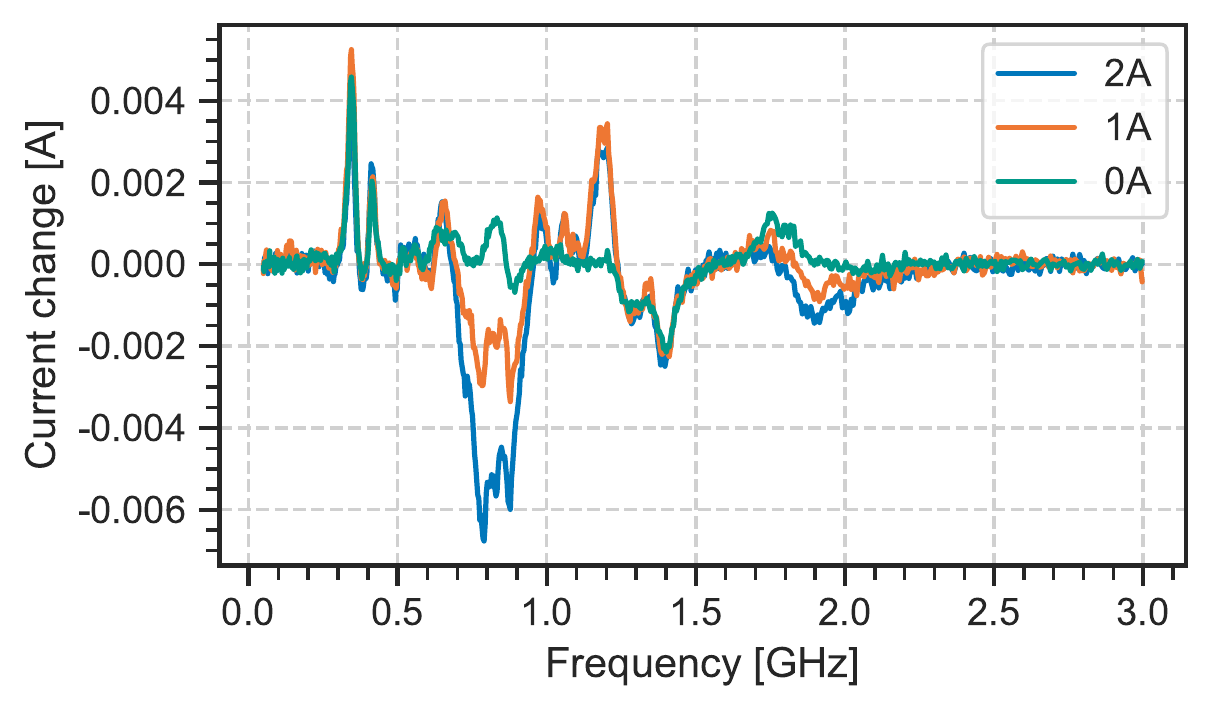}
        \caption{Current sensor INA260 at various real currents.}
        \label{fig:cs_id4}
    \end{subfigure}\\
    \caption{Change in the current measured by the remaining automotive current sensors.}
    \label{fig:cs_extra}
\end{figure*}

Out of the 4 current sensors, we show one in the paper, and one further sensor did not respond to the attack.
Results from the remaining two can be seen in Figure~\ref{fig:cs_extra}.

\subsection{Constant current supplies}

\label{app:plots_cc}
\begin{figure}[t]
    \centering
    \includegraphics[width=\columnwidth]{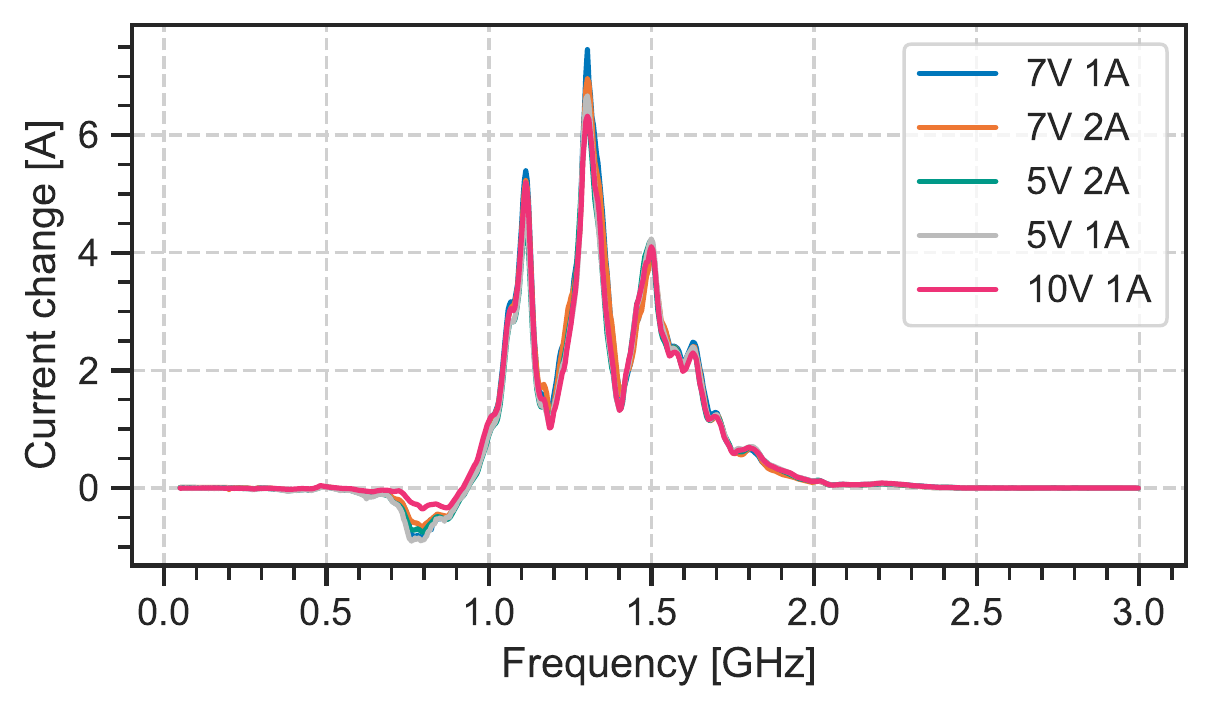}
    \caption{Frequency response of DC-DC \#6 in constant current mode at various applied CV loads and current limits.}
    \label{fig:dcdc-cc-loads}

    \label{fig:cc_extra}
\end{figure}

We investigated constant current supplies, and conclude that many of them are extremely vulnerable. 
We also investigated how the current limit and load voltage impact the attack against current sensors.
We performed this evaluation for device \#6, as it was the most stable current source we had purchased.
We show these results on Figure~\ref{fig:cc_extra}, and conclude that the absolute current offset is constant regardless of the current limit and load.

\subsection{Chargers}

\label{app:plots_chg}
\begin{figure}[t]
    \centering
    \includegraphics[width=\columnwidth]{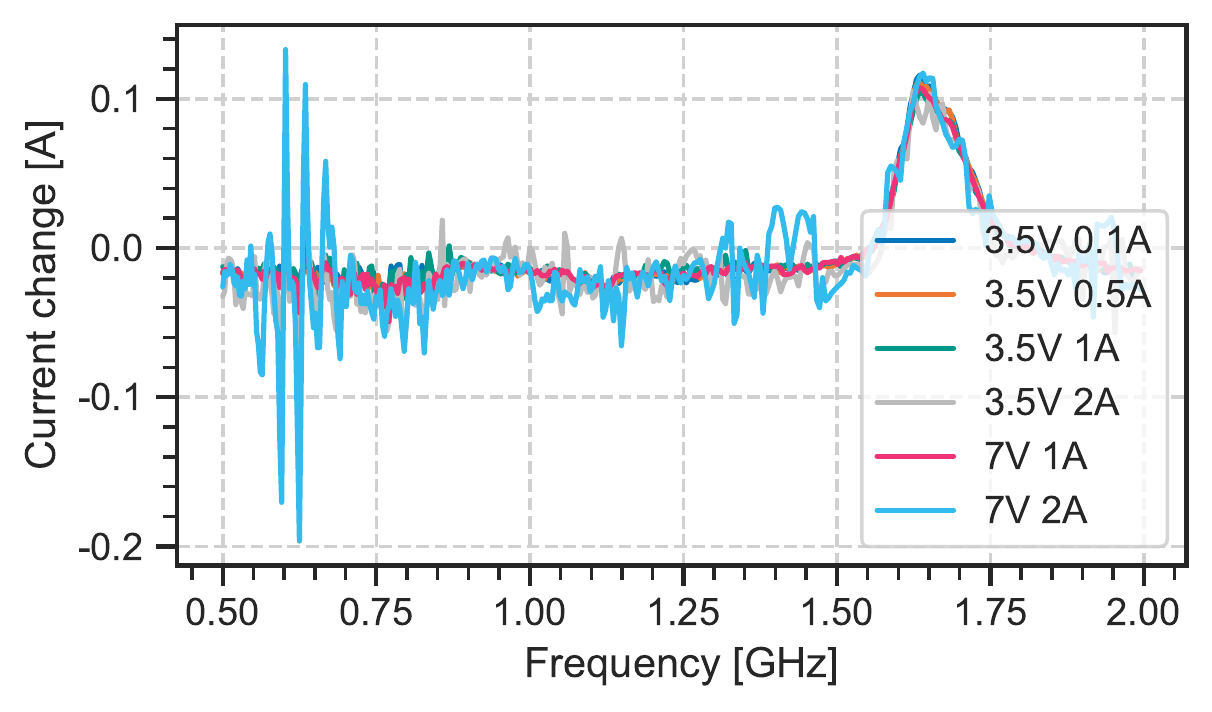}
    \caption{Frequency response of the C240 Duo in CC mode at various loads.}
    \label{fig:chg_id2}

    \label{fig:chg_extra}
\end{figure}

We tested two chargers, the C240 Duo and the SkyRC e4.
We found that the current attack works well on both of them, and achieves a constant offset regardless of the set current.
In the paper we present this plot for the SkyRC e4, and in Figure~\ref{fig:chg_id2} we present the results for the C240 Duo.

\end{document}